\documentclass[twocolumn]{aastex631}

\usepackage{bm}
\usepackage{amsmath}
\usepackage{graphicx}       
\usepackage{multirow}
\usepackage{soul}

\begin{document}

\title{Robust Evidence for the Breakdown of Standard Gravity at Low Acceleration from Statistically Pure Binaries Free of Hidden Companions}

\correspondingauthor{Kyu-Hyun Chae}
\email{chae@sejong.ac.kr, kyuhyunchae@gmail.com}

\author[0000-0002-6016-2736]{Kyu-Hyun Chae}
\affiliation{Department of Physics and Astronomy, Sejong University, 209 Neungdong-ro Gwangjin-gu, Seoul 05006, Republic of Korea}



\begin{abstract}
  It is found that Gaia DR3 binary stars selected with stringent requirements on astrometric measurements and radial velocities naturally satisfy Newtonian dynamics without hidden close companions when projected separation $s \lesssim 2$~kau, showing that pure binaries can be selected. It is then found that pure binaries selected with the same criteria show a systematic deviation from the Newtonian expectation when $s \gtrsim 2$~kau. When both proper motions and parallaxes are required to have precision better than 0.005 and radial velocities better than 0.2, I obtain 2,463 statistically pure binaries within a `clean' $G$-band absolute magnitude range. From this sample, I obtain an observed to Newtonian predicted kinematic acceleration ratio of $\gamma_g=g_{\rm{obs}}/g_{\rm{pred}}=1.49^{+0.21}_{-0.19}$ for acceleration $\lesssim 10^{-10}$~m~s$^{-2}$, in excellent agreement with $1.49\pm 0.07$ for a much larger general sample with the amount of hidden close companions self-calibrated. I also investigate the radial profile of stacked sky-projected relative velocities without a deprojection to the 3D space. The observed profile matches the Newtonian predicted profile for $s \lesssim 2$~kau without any free parameters but shows a clear deviation at a larger separation with a significance of $\approx 5.0\sigma$. The projected velocity boost factor for $s\gtrsim 5$~kau is measured to be $\gamma_{v_p} = 1.20\pm 0.06$ (stat) $\pm 0.05$ (sys) matching $\sqrt{\gamma_g}$. Finally, for a small sample of 40 binaries with exceptionally precise radial velocities (fractional error $<0.005$) the directly measured relative velocities in the 3D space also show a boost at larger separations. These results robustly confirm the recently reported gravitational anomaly at low acceleration for a general sample.
\end{abstract}



\section{Introduction} \label{sec:intro}

 {Wide binaries (widely-separated, long-period, gravitationally-bound binary stars) provide crucial testbeds for probing gravitational dynamics in the low-acceleration regime (e.g., \citealt{hernandez2012,banik2018,pittordis2018,banik2019,pittordis2019,hernandez2022}).} A couple of recent studies by \cite{chae2023} and \cite{hernandez2023} of wide binary stars selected from Gaia data release 3 \citep[DR3;][]{dr3} have reported a gravitational anomaly at low acceleration $\la 10^{-9}$~m~s$^{-2}$, or for a sky-projected separation $s \ga 2$~kau (kilo astronomical unit) for typical binaries with total masses of $\sim (0.5-2){\rm{M}}_\odot$. This gravitational anomaly implies a low-acceleration breakdown of both Newtonian dynamics and general relativity and so has immense implications for astrophysics, cosmology, and fundamental physics. Thus, one cannot overemphasize the importance of confirming the claimed anomaly from as many independent studies as possible.

\cite{chae2023} considered wide binaries selected from \cite{elbadry2021} that are statistically free of both chance-alignment cases and resolved ($>1''$) triples and higher-order multiples. Because of the initial selection, additional quality cuts, and the availability of dust extinction correction, \cite{chae2023} used only up to $26,615$ wide binaries within 200~pc. \cite{chae2023} then self-calibrated the occurrence rate ($f_{\rm{multi}}$) of triples and higher-order multiples hosting hidden (i.e.\ unresolved) close companions by requiring that binaries must satisfy Newtonian dynamics at a close enough separation, or at a high enough acceleration $\ga 10^{-8}$~m~s$^{-2}$, as predicted by all currently available plausible theories including modified Newtonian dynamics \citep[MOND;][]{milgrom1983}. \cite{chae2023} also paid a particular attention to projection effects and employed a Monte Carlo (MC) method to deproject measured sky-projected relative velocities $v_p$ to the three-dimensional (3D) space physical velocities $v$, and compared a kinematic acceleration $v^2/r$ with the corresponding Newtonian prediction. \cite{chae2023} obtained up to a $10\sigma$ significance for the gravitational anomaly based on MC analyses. Moreover, the magnitude and trend of the anomaly matched well the prediction of MOND-type modified gravity such as AQUAL \citep{bekenstein1984} and QUMOND \citep{milgrom2010}  {under the external field effect (EFE) of the Milky Way}. 

\cite{hernandez2023} took a different approach that tried to remove all cases of both chance alignments and hierarchical multiples. Because \cite{hernandez2023} applied various strict cuts, his final sample includes only 450 pure binaries in the distance range $d<125$~pc or $125<d<170$~pc. \cite{hernandez2023} calculated the dispersion of one-dimensional velocity components on the plane of the sky and compared it with the Newtonian prediction by \cite{jiang2010}. \cite{hernandez2023} checked that small-separation ($s\la 2$~kau) systems matched the Newtonian prediction indicating that kinematic contaminants are indeed negligible and Newtonian dynamics holds in the high-acceleration regime. Then, he found that the observed sky-projected velocities systematically deviated from the Newtonian expectation at large-separation ($s\ga 2$~kau) systems. Because the sample size was small and the observed kinematics of the binaries was compared with simulations for other binaries, \cite{hernandez2023} did not quantify a statistical significance of the anomaly seen at $s\ga 2$~kau. Nevertheless, the final result from \cite{hernandez2023} agreed with that of \cite{chae2023}. 

Unlike \cite{chae2023} and \cite{hernandez2023}, another recent study by \cite{pittordis2023} based on a Gaia database considered only low-acceleration binaries and thus could not calibrate $f_{\rm{multi}}$ among their wide binaries. Their analysis was also compounded by their inclusion of chance-alignment cases. \cite{pittordis2023} erroneously concluded that Newtonian dynamics matched their low-acceleration data with their ``fitted'' $f_{\rm{multi}}$ (without a proper calibration). As \cite{chae2023} demonstrated, \cite{pittordis2023} conclusion is not surprising because the dynamics uncovered by \cite{chae2023} is pseudo-Newtonian with a rescaling of Newton's constant $G\rightarrow 1.4 G$ and this boost can be canceled by a higher $f_{\rm{multi}}$ (without a proper calibration).

 {Just recently, \cite{banik2023} argued that Newtonian dynamics was preferred over MOND (using specifically the QUMOND model) based on a method similar to \cite{pittordis2023}. They did not include the Newtonian regime ($\la 2$~kau) data to calibrate $f_{\rm{multi}}$ but claimed that gravity and $f_{\rm{multi}}$ could be simultaneously constrained based only on data from the Newton-MOND transition and MOND regimes (see below). They obtained a high value of $f_{\rm{multi}}\approx 0.70$ for their preferred Newtonian model although their sample included only binaries passing a strict cut on relative velocities. Their value of $f_{\rm{multi}}\approx 0.70$ is unlikely high compared with the observed range $0.3 \la f_{\rm{multi}} \la 0.5$ (e.g., \citealt{raghavan2010,riddle2015,moe2017}) even for general binary samples without kinematic cuts. Moreover, I note that binaries selected by strict kinematic cuts have significantly lower $f_{\rm{multi}}$ than that for a general sample. I will further discuss their sample selection, analyses, and results at relevant places (see in particular appendices). }

\begin{figure}
  \centering
  \includegraphics[width=1.\linewidth]{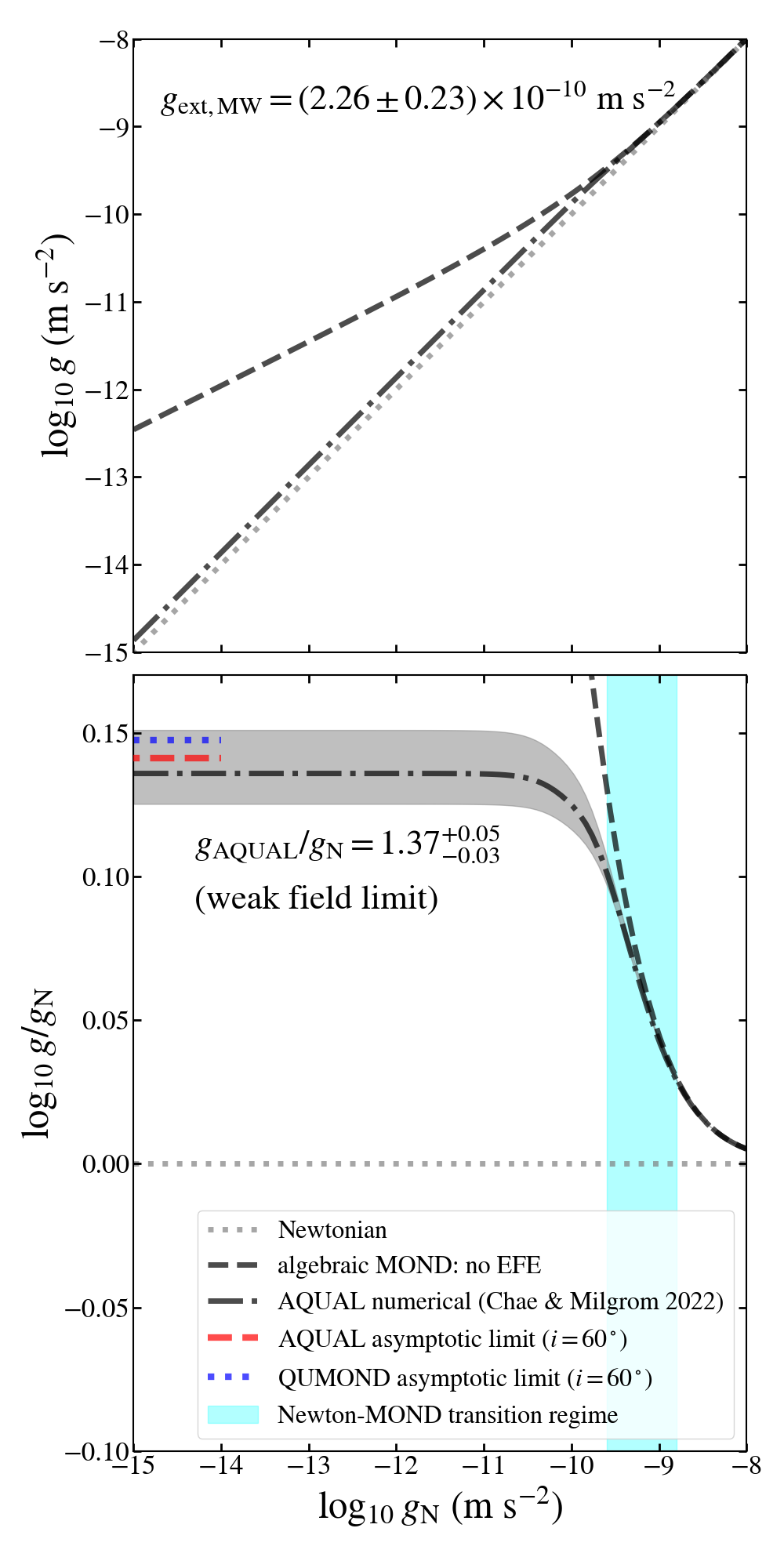}
    \vspace{-0.5truecm}
    \caption{\small 
    {This figure shows an AQUAL numerical \citep{chae2022a} prediction on the internal gravitational field for circular orbits under the external field of the Milky Way as estimated in \cite{chae2023}. The numerical acceleration matches well the analytic asymptotic value when the external field is inclined at an average angle of $60^\circ$ from the orbital axis. Internal dynamics is expected to switch from the Newtonian regime to the MOND regime over the narrow acceleration range indicated by the cyan-colored band.}
    } 
   \label{EFE_MW}
\end{figure}

\begin{figure}
  \centering
  \includegraphics[width=1.\linewidth]{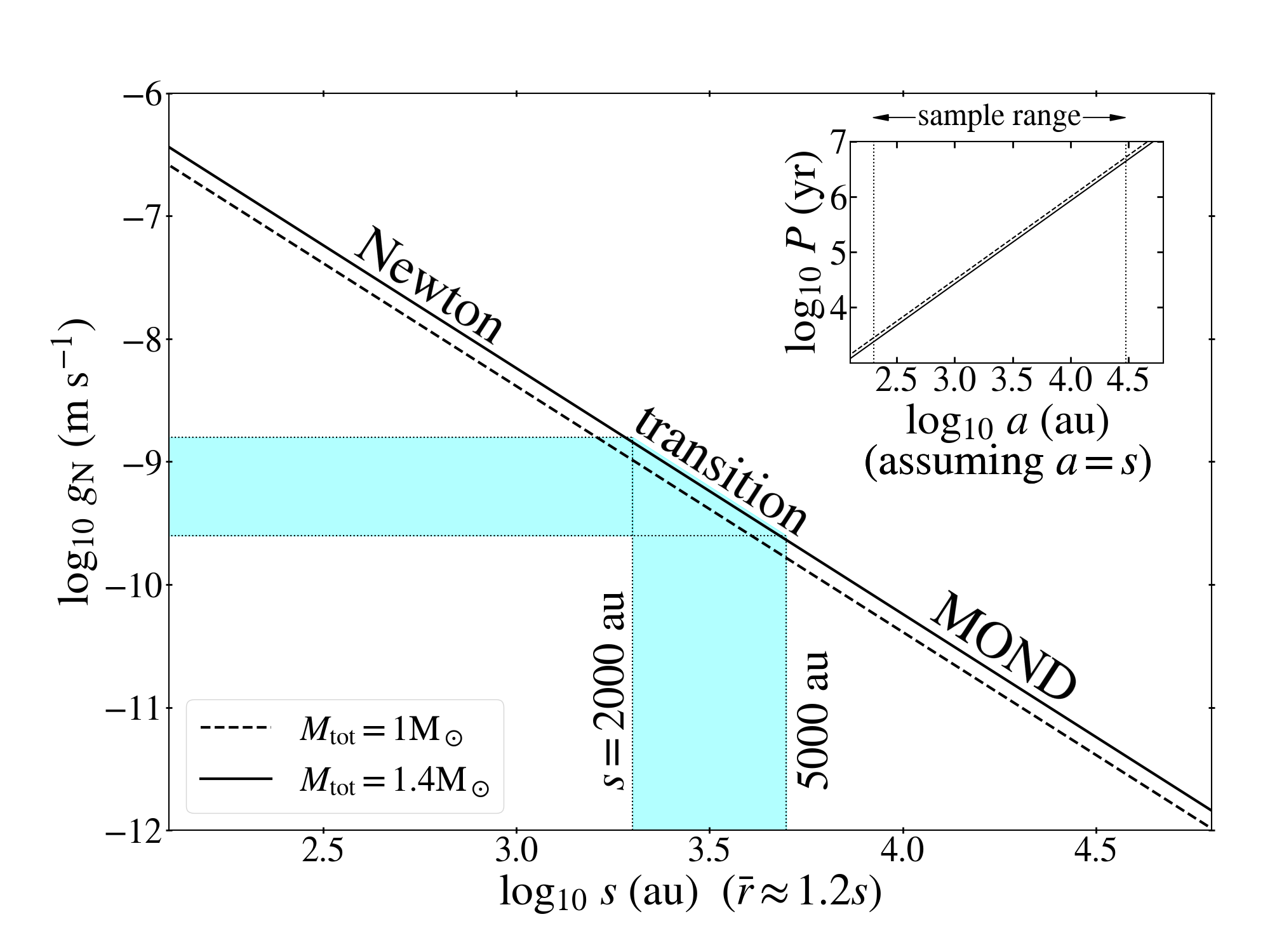}
    \vspace{-0.5truecm}
    \caption{\small 
    {This figure shows the range of sky-projected separation $s$ corresponding to the Newton-MOND transition range in acceleration shown in Figure~\ref{EFE_MW}. The inset shows the Keplerian prediction on the range of orbital periods for Gaia wide binaries used in \cite{chae2023} and this study.}
    } 
   \label{transition}
\end{figure}

 Figure~\ref{EFE_MW} shows an AQUAL numerical \citep{chae2022a} prediction on the radial acceleration for circular orbits under the EFE of the Milky Way that approximately matches the AQUAL analytic asymptotic limit at an average inclination of the external field. Due to the strong external field of the Milky Way, internal dynamics is expected to switch from the Newtonian regime to the pseudo-Newtonian regime with a boosted Newton's constant. As shown in Figure~\ref{EFE_MW}, most of the transition is expected to occur abruptly in the narrow acceleration range of $-9.6\la \log_{10}(g_{\rm{N}}/\text{m s}^{-2}) \la -8.8$. As shown in Figure~\ref{transition}, for typical Gaia wide binaries with a total mass of $\approx 1.4 M_\odot$ the transition acceleration range corresponds approximately to the sky-projected separation range of $2\la s \la 5$~kau (kilo astronomical units). This MOND prediction was supported by the two published analyses \citep{chae2023,hernandez2023} and is intended to be further tested in this study.  

Here I consider a new analysis that is complementary to  \cite{chae2023} and \cite{hernandez2023}) and can provide a robust test of gravity. The analysis of \cite{chae2023} involved a complex chain of steps with various observational inputs for a general sample and obtained a maximal statistical power. The complexity arises largely from modeling the kinematic effects of hidden close companions with currently available observational inputs. As $f_{\rm{multi}}\rightarrow 0$, the complexity is gone and thus any possibilities of systematic errors involving close companions can be removed at the cost of losing statistical power. The question is whether a statistically significant sample of $f_{\rm{multi}} = 0$ can be selected in a systematic and verifiable way so that gravitational anomaly can be tested robustsly and with a sufficient statistical power.

\cite{hernandez2023} obtained a sample  {that was supposed to be largely free of hierarchical systems} but his analysis was limited in two ways. \cite{hernandez2023} did not carry out a Monte Carlo simulation or any other statistical procedure for his sample to quantify a statistical significance of the seen anomaly (although a comparison with an independent simulation by \cite{jiang2010} was made). Also, each sample defined by \cite{hernandez2023} seems too small with just 450 binaries.

Here I obtain a much larger sample of  {2,463} wide binaries of $f_{\rm{multi}} = 0$ in a systematic and verifiable way. Then, I test gravity in an acceleration plane with the algorithm developed in \cite{chae2023} with $f_{\rm{multi}} = 0$. More importantly, I investigate stacked velocity profiles with a MC simulation to do a quantitative statistical test. In this way, the present work will complement both \cite{chae2023} and \cite{hernandez2023}. I will show that the results from pure binaries agree excellently with those of \cite{chae2023} reaffirming the validity of the procedure and conclusion of \cite{chae2023}.

The structure of this paper is as follows. Section~\ref{sec:data} describes how a sample of pure binaries can be selected in a systematic way. Section~\ref{sec:method} describes a Monte Carlo modeling of pure binary stars. Section~\ref{sec:result} presents the results. Section~\ref{sec:discussion} discusses any possible source of systematic errors. In Section~\ref{sec:conclusion}, I offer the conclusion and discuss future works.  In Appendix~\ref{sec:correction}, I describe a correction to \cite{chae2023} and revise the representative results.  {In Appendix~\ref{sec:kincut}, I consider some kinematic quality cuts on general samples of wide binaries and their effects on inference on gravity. } Python scripts used for this work and the sample of pure binaries can be accessed at Zenodo: doi:10.5281/zenodo.8416435.

\section{A systematic selection of pure binaries} \label{sec:data}

Following \cite{chae2023}, I work with the catalog of one million candidate binaries derived by \cite{elbadry2021} from Gaia DR3 astrometric measurements. This catalog provides estimated values of chance-alignment probability ($\mathcal{R}$) so that chance-alignment cases can be effectively excluded. The catalog also excludes triples and higher-order multiples whose components are all resolved by $>1''$. Thus, by requiring $\mathcal{R}<0.01$ (or something similar), one can choose binaries that may include only unresolved close companions.

Specifically, I consider  {26,615} binaries with $\mathcal{R}<0.01$ within 200~pc from the Sun whose components have sky-projected separation ($s$) in the range $0.2<s<30$ kau and have absolute magnitudes in the `clean range' $4<M_G<14$ defined by \cite{chae2023} where $M_G$ is the dust-extinction-corrected absolute magnitude in the Gaia $G$ band. As \cite{chae2023} showed, samples of binaries selected by a precision threshold of $0.01$ or $0.005$ imposed on the measured proper motions (PMs) require $f_{\rm{multi}} \ga 0.3$ to statistically satisfy Newtonian dynamics at high enough accelerations $\ga 10^{-8}$~m~s$^{-2}$.

Because binaries must satisfy Newtonian dynamics at high enough accelerations, binaries selected in a systematic way may be regarded as a sample of pure binaries statistically free of hidden close companions if they require $f_{\rm{multi}}=0$ to match Newtonian dynamics at accelerations $\ga 10^{-8}$~m~s$^{-2}$. For this procedure to be valid masses of individual stars must be reliably known. Fortunately, \cite{chae2023} provides a couple of reliable magnitude-mass relations (see figure~7 and table~1 of \cite{chae2023}) in the Gaia $G$ band. In search of a sample of pure binaries satisfying $f_{\rm{multi}}=0$, various selection criteria have been tried guided by observational and simulation studies (e.g., \citealt{belokurov2020,penoyre2022}).

If a close undetected companion is present, it can have various effects. First, the image may not be well modeled so that Gaia derived {\tt ruwe} values may be significantly larger than 1. Second, because the close companion induces additional motions, the measured uncertainties of parallaxes and proper motions (PMs) will become larger. Third, the additional motions will also increase the measurement uncertanties of radial velocities.

It turns out that a sample of pure binaries can be obtained by requirements on both astrometric measurements and radial velocities. The requirements can be specified as follows. Throughout the brighter (more massive) star is referred to as component~A while the fainter star component~B. 

\begin{figure*}
  \centering
  \includegraphics[width=0.9\linewidth]{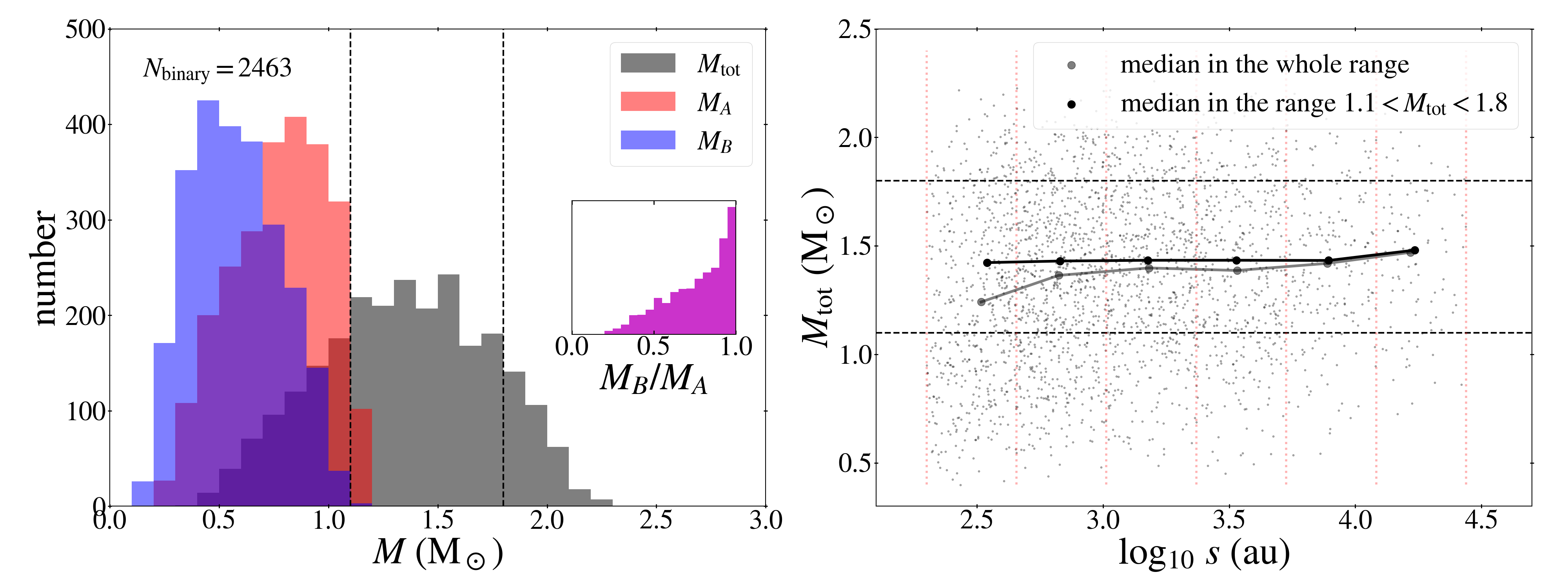}
    \vspace{-0.2truecm}
    \caption{\small 
    The left panel shows distributions of individual masses and total masses in  {2,463} pure binaries of the main sample defined in the text. The dashed black vertical lines define a narrow mass range of total masses $1.1<M_{\rm{tot}}/{\rm{M}}_\odot<1.8$. The right panel shows the mean masses in the  {6} bins defined by sky-projected separation $s$ as indicated by vertical red dashed lines. These  {6} bins will be used in the analyses of sky-projected velocities.
    } 
   \label{massdist}
\end{figure*}

\begin{enumerate}

\item Reported values of {\tt ruwe} for both components are smaller than 1.2.
  
\item Both PM components and parallaxes (thus distances) have relative (i.e.\ fractional) measurement errors smaller than $\varepsilon$ with $\varepsilon \le 0.005$.

\item Distances of two components agree within $2\sigma$ uncertainties and a maximum possible difference from the elliptical orbit $\Delta d_{\rm{orbit}}^{\rm{max}}$, i.e., 
  \begin{equation}
    \left|d_{\rm{A}} - d_{\rm{B}}\right| < \sqrt{4(\sigma_{d_{\rm{A}}}^2+\sigma_{d_{\rm{B}}}^2)+(\Delta d_{\rm{orbit}}^{\rm{max}})^2},
    \label{eq:deld}
    \end{equation}
where $\Delta d_{\rm{orbit}}^{\rm{max}}=6s$  from a 99\% statistical limit from random inclinations and orbital phases for elliptical orbits of observational eccentricities.

\item Radial velocities of both components have relative measurement errors smaller than 0.2. In other words, we take only binaries whose components have radial velocities with $S/N > 5$.

\item Radial velocities of two components agree within $2\sigma$ uncertainties and a maximum possible difference from the elliptical orbit $\Delta v_{r,\rm{orbit}}^{\rm{max}}$, i.e.,
  \begin{equation}
    \left|v_{r,\rm{A}}-v_{r,\rm{B}}\right| < \sqrt{4(\sigma_{v_{r,\rm{A}}}^2+\sigma_{v_{r,\rm{B}}}^2)+(\Delta v_{r,\rm{orbit}}^{\rm{max}})^2}
    \label{eq:delvr}
  \end{equation}
  with
  \begin{equation}
   \Delta v_{r,\rm{orbit}}^{\rm{max}} = 0.9419{\text{ km s}^{-1}}\sqrt{\frac{M_{\rm{tot}}}{s}}\times 1.3\times 1.2, 
    \label{eq:vmax}
  \end{equation}
  where $M_{\rm{tot}}$ is the total mass of the binary system in units of solar mass (M$_\odot$) and $s$ is the sky-projected separation between the two components. The factor 1.3 represents a maximum possible value (see Section~\ref{sec:method} below) arising from random inclinations and orbital phases for elliptical orbits of observational eccentricities.  The last factor 1.2 allows for a possible boost of velocity in MOND-type modified gravity theories so as not to preclude such theories, though it is practically not important because other uncertainties are larger.
  
\end{enumerate}

In the above, the third and fifth requirements gaurantee that two stars form a true binary system. When $\varepsilon = 0.005$ is used, the total number of pure binary systems is $N_{\rm{tot}}=2463$. The distribution of masses of the selected binaries can be found in Figure~\ref{massdist}. The total masses are in the range $0.5\la M_{\rm{tot}}/{\rm{M}}_\odot \la 2.1$ with a mass ratio $M_B/M_A\ge 0.5$ for 88\% of the systems. The mean (median) mass for the entire sample is $1.35$ ($1.36$)~M$_\odot$, and the binned mean varies only mildly with $s$ as the right panel of Figure~\ref{massdist} shows. I also consider a subsample within a narrow range of total mass $1.1 < M_{\rm{tot}}/{\rm{M}}_\odot < 1.8$. This subsample has a mean (median) mass of $1.44$ ($1.43$)~M$_\odot$ and its binned mean does not vary with $s$.

Since the mean mass varies little or not at all with $s$ in the entire sample or the subsample, it is possible to investigate a stacked radial velocity profile. This is important because a statistical analysis of velocity profiles will be a main part of this work.  

For the observed right ascension ($\alpha$) and declination ($\delta$) components of the PMs in a binary, $(\mu_{\alpha,A}^\ast, \mu_{\delta,A})$ and $(\mu_{\alpha,B}^\ast, \mu_{\delta,B}),$\footnote{Here $\mu_\alpha^\ast\equiv \mu_\alpha \cos\delta$ for PM component $\mu_\alpha$.} along with the accurately and precisely measured distances $d_A$ and $d_B$, the magnitude of of the plane-of-sky relative velocity $v_p$ is given by
\begin{equation}
 v_p = \left[(\mu_{\alpha,A}^\ast d_A - \mu_{\alpha,B}^\ast d_B)^2 + (\mu_{\delta,A}d_A - \mu_{\delta,B}d_B )^2\right]^{1/2}.
  \label{eq:vp_ob_exact}
\end{equation}
For Equation~(\ref{eq:vp_ob_exact}) to be used for actual data, the precision of $d_A$ and $d_B$ must be extremely good to prevent spurious boost of $v_p$ in some systems caused by random measurement errors. Thus, in practice it is more accurate to use
\begin{equation}
  v_p =  4.7404\times 10^{-3}\text{ km s}^{-1}\times \Delta\mu \times  d
  \label{eq:vp_ob}
\end{equation}
where $d$ is a representative distance in pc to the binary system, and
\begin{equation}
 \Delta\mu = \left[(\mu_{\alpha,A}^\ast - \mu_{\alpha,B}^\ast )^2 + (\mu_{\delta,A} - \mu_{\delta,B} )^2\right]^{1/2},
  \label{eq:PM}
\end{equation}
with all PM values given units of mas~yr$^{-1}$. For $d$ I take an error-weighted mean ($d_M$) of $d_A$ and $d_B$.

\begin{figure}
  \centering
  \includegraphics[width=1.\linewidth]{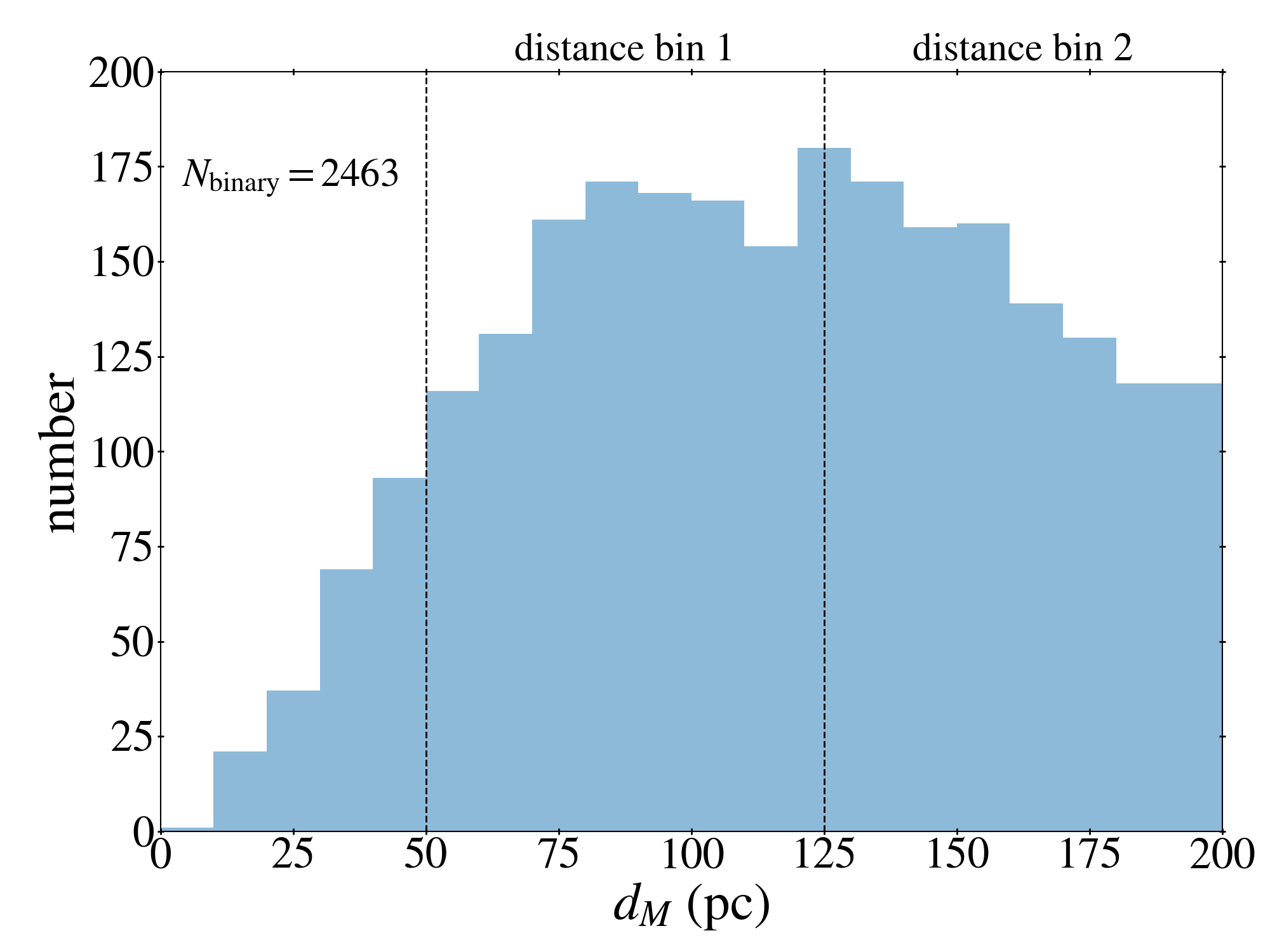}
    \vspace{-0.2truecm}
    \caption{\small 
     {This figure shows the distribution of $d_M$ (the error-weighted mean distance) for the sample of statistically pure binaries shown in Figure~\ref{massdist}. Two distance bins are indicated by the vertical dashed lines. These bins will be used to probe any possible systematic effect of distances.}
    } 
   \label{dist}
\end{figure}

Tests with the Gaia sample show that velocities estimated with Equations~(\ref{eq:vp_ob_exact}) and (\ref{eq:vp_ob}) are statistically equivalent only when the precision of distances is better than $\varepsilon \approx 0.002$. In this work Equation~(\ref{eq:vp_ob}) will be used because up to $\varepsilon \approx 0.005$ is considered. Note also that because the distance range is $9\la d/\text{pc} \la 200$ (Figure~\ref{dist}) and $8\times 10^{-6}\la s/d\la 2.8\times 10^{-3}$ with a median of $5\times 10^{-5}$, it is sufficient to assume a plane geometry for the sky region of a binary system.

 {The uncertainty of the PM magnitude (Equations~(\ref{eq:PM})) is estimated following \cite{elbadry2021} as}
\begin{equation}
  \begin{array}{ccl}
    \sigma_{\Delta\mu} & = & \left[(\sigma_{\mu_{\alpha,A}^\ast}^2 + \sigma_{\mu_{\alpha,B}^\ast}^2 )(\Delta\mu_\alpha)^2\right. \\
      &  & \left. + (\sigma_{\mu_{\delta,A}}^2 + \sigma_{\mu_{\delta,B}}^2 )(\Delta\mu_\delta)^2 \right]^{1/2}/\Delta\mu, \\
    \end{array}
  \label{eq:PMerr}
\end{equation}
 { where }
\begin{equation}
  \begin{array}{ccl}
  (\Delta\mu_\alpha)^2 & = & (\mu_{\alpha,A}^\ast - \mu_{\alpha,B}^\ast )^2, \\
  (\Delta\mu_\delta)^2 & = & (\mu_{\delta,A} - \mu_{\delta,B})^2. \\
    \end{array}
  \label{eq:PMcomp}
\end{equation}
 {The uncertainty of the sky-projected velocity is given by}
\begin{equation}
  \sigma_{v_p} =  4.7404\times 10^{-3}\text{ km s}^{-1}\times \sigma_{\Delta\mu} \times  d.
  \label{eq:sigvp_ob}
\end{equation}
 {The normalized velocity parameter $\tilde{v}$ \citep{banik2018} and its uncertainty are given by}
\begin{equation}
  \begin{array}{ccl}
    \tilde{v} & \equiv & v_p/v_c, \\
    \sigma_{\tilde{v}} & \equiv & \tilde{v}\sqrt{\left(\frac{\sigma_{v_p}}{v_p}\right)^2+\left(\frac{\sigma_{v_c}}{v_c}\right)^2}, \\
    \end{array}
  \label{eq:vtilde}
\end{equation}
 {where $v_c\equiv\sqrt{GM_{\rm{tot}}/s}$ is the Newtonian circular velocity defined at the projected separation $s$ \citep{banik2018} and I take $\sigma_{v_c}/v_c=0.05$ assuming a total mass uncertainty of 10\%. The uncertainties of $v_p$ and $\tilde{v}$ are introduced here as additional means to check/control data quality. }

\begin{figure}
  \centering
  \includegraphics[width=0.8\linewidth]{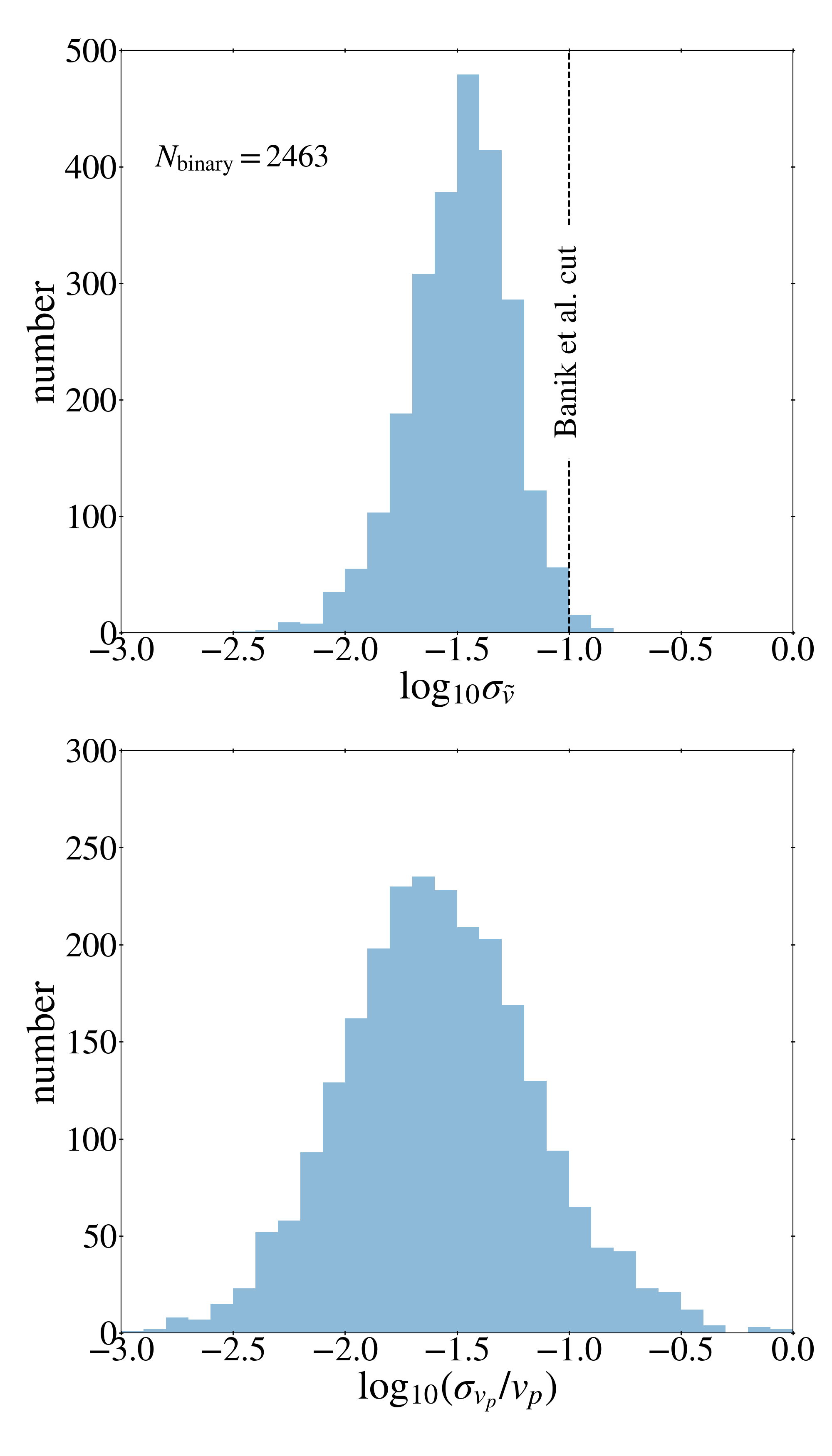}
    \vspace{-0.2truecm}
    \caption{\small 
      {The upper panel shows the distribution of $\sigma_{\tilde{v}}$ (Equation~(\ref{eq:vtilde})) while the bottom panel shows the distribution of the fractional uncertainty of the sky-projected velocity $\sigma_{v_p}/v_p$ for the main sample of 2,463 pure binaries. The upper panel indicates the cut similar to that suggested by \cite{banik2023}. Note that virtually all pure binaries satisfy the \cite{banik2023} cut and $S/N>2$ for the sky-projected velocity.}
    } 
   \label{verr}
\end{figure}

 {Figure~\ref{verr} shows the distributions of the estimated uncertainties of $v_p$ and $\tilde{v}$ for the pure binaries of the main sample shown in Figure~\ref{massdist}. Virtually all pure binaries selected with strict astrometric and kinematic criteria have good signal-to-noise ($S/N\ga 3$) with a median of about 40 for $v_p$. \cite{banik2023} advocate a cut based on $\sigma_{\tilde{v}}<0.1\max(1,\tilde{v}/2)$. Figure~\ref{verr} shows that this cut is already satisfied by virtually all pure binaries. Thus, for the pure binary sample I will not consider any artificial cut using either $\sigma_{v_p}/v_p$ or $\sigma_{\tilde{v}}$. In Appendix~\ref{sec:kincut}, I will discuss the effects of cuts in general samples.}
    
Gaia DR3 radial velocities typically have much less precision than PMs. Thus, most radial velocities cannot be used to measure the relative radial velocity $v_r$ between the two components because large random errors in individual radial velocities can create spurious boosts in many systems. However, for exceptionally precise radial velocities with the measurement precision comparable to that of $\Delta\mu$ (Equation~(\ref{eq:PM})), two relative velocities $v_p$ and $v_r$ can be combined to reliably estimate the relative physical velocity between the two stars
\begin{equation}
  v =  \sqrt{v_p^2 + v_r^2}.
  \label{eq:v_ob}
\end{equation}
 {Considering that all PM components have relative errors $<\varepsilon$, the relative error of $v_{p,i}$ ($i=\alpha,\delta$)\footnote{For most confirmed binary systems the two stars can be regarded at the same distance, so the distance may be treated as a constant.} is $<\sqrt{2}\varepsilon$ for $\varepsilon=0.005$. To require that the relative error of $v_r$ is comparable with that of $v_{p,i}$, I require relative errors of individual radial velocities $<0.005$.} Finally, I note that gravitational redshifts \citep{elbadry2022} from the surface gravities of the stars are irrelevant for the stars used in this work because stellar mass-to-radius ratio $M/R$ varies little for the mass range $0.1\la M/\text{M}_\odot\la 1$ \citep{demory2009}.

\begin{figure*}
  \centering
  \includegraphics[width=0.8\linewidth]{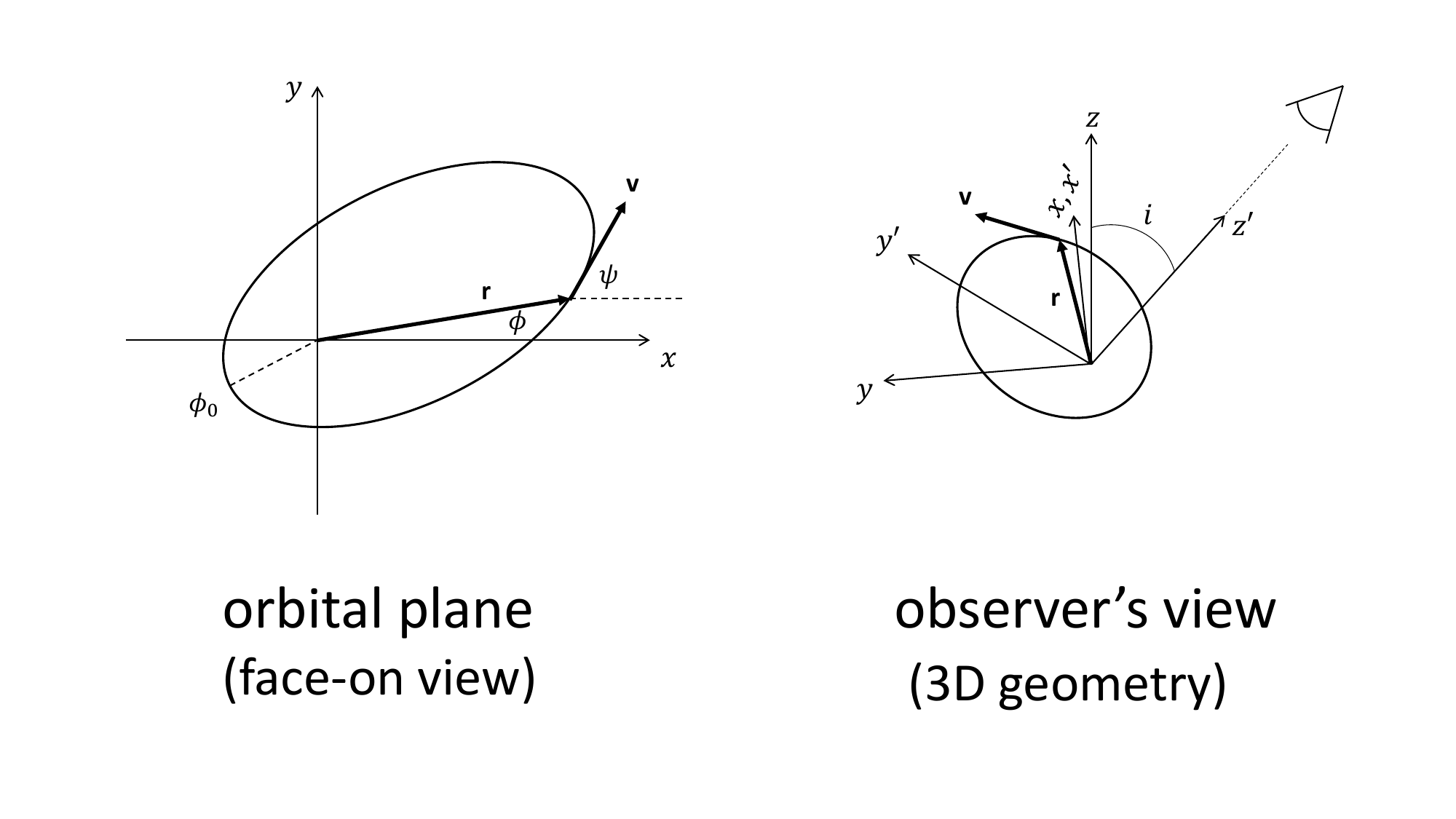}
    \vspace{-0.2truecm}
    \caption{\small 
    (Adapted from \cite{chae2023}) The left panel shows a one-particle equivalent description of orbital motions of the two stars in a binary system. The right panel defines the observer's viewpoint at an inclination $i$. 
    } 
   \label{orbit}
\end{figure*} 

\section{A Monte Carlo method of testing gravity with stacked velocity profiles of pure binaries} \label{sec:method}

Testing gravity in an acceleration plane with pure binaries will be done using the algorithm of \cite{chae2023} with $f_{\rm{multi}}=0$. Here I describe a Monte Carlo method of testing gravity with stacked velocity profiles of pure binaries. The description of elliptical orbits will be the same as that of \cite{chae2023}. The pertinent question is how to predict the sky-projected relative velocity $v_p(s)$ and the physical relative velocity $v(s)$ between the two stars with a sky-projected separation $s$ that can be compared with the observed velocities, Equations~(\ref{eq:vp_ob}) and (\ref{eq:v_ob}).

Figure~\ref{orbit} shows an equivalent one-body description of the elliptical orbit of binary dynamics taken from \cite{chae2023}. The orbit is described in the plane polar coordinates $(r,\phi)$ by the equation
  \begin{equation}
   r = \frac{a(1-e^2)}{1+e\cos(\phi-\phi_0)}, 
    \label{eq:orbit}
  \end{equation}
where $e$ is the eccentricity, $a$ is the semi-major axis, and $\phi_0$ is the longitude of the periastron. 
  
In Newtonian dynamics, the magnitude of the relative physical velocity between the two stars is given by 
  \begin{equation}
    v(r) = \sqrt{\frac{GM_{\rm{tot}}}{r} \left(2- \frac{r}{a} \right)}.
    \label{eq:vN}
  \end{equation}
Physical separation $r$ is related to the sky-projected separation $s$ by
  \begin{equation}
    s = r \sqrt{1 - \sin^2i \sin^2\phi},
    \label{eq:s_r}
  \end{equation}
where $i$ is the inclination and $\phi$ is the azimuthal angle of the physical separation vector on the orbital plane as shown in Figure~\ref{orbit}.
  
Combining Equations~(\ref{eq:orbit}), (\ref{eq:vN}), and (\ref{eq:s_r}), we can express the magnitude of the relative physical velocity as a function $s$,
  \begin{equation}
      \begin{array}{lll}
    v(s) & = &  0.9419\text{ km s}^{-1}\sqrt{\frac{M_{\rm{tot}}/\text{M}_\odot}{s/\text{kau}}}\times \\
     &  &  \sqrt{\sqrt{1 - \sin^2i \sin^2\phi} \left( 2 - \frac{1-e^2}{1+e\cos(\phi-\phi_0)} \right)}. \\
      \end{array}
    \label{eq:vN_s}
  \end{equation}
The magnitude of the sky-projected velocity to the observer is given by
  \begin{equation}
    v_p(s) = v(s) \sqrt{1 - \sin^2i \sin^2\psi},
    \label{eq:vpN_s}
  \end{equation}
where 
 \begin{equation}
  \psi = \tan^{-1} \left( - \frac{\cos\phi+e\cos\phi_0}{\sin\phi+e\sin\phi_0} \right) +\pi.
  \label{eq:psi}
 \end{equation}
 {(I note that the factor $\pi$ is physically irrelevant and added to match the definition given in Figure~\ref{orbit} exactly.)}
 
For the observed set of $(M_{\rm{tot}},s)$, one MC realization of Newtonian velocities of Equations~(\ref{eq:vN_s}) and (\ref{eq:vpN_s}) follow from MC realizations of $\phi_0$, $\phi$, $i$ and $e$. Because possible ranges of these parameters are broad, the predictions for one binary system cannot be meaningfully compared with the observed velocities to test gravity. However, if a number of binary systems are considered simultaneously, the individual random fluctuations are averaged out and thus the mean of the predictions can be meaningfully compared with the mean of the observed velocities. Moreover, if MC realizations are repeated many times, on can derive the probability distribution of the mean in a sample and thus estimate its statistical uncertainty. This procedure allows one to test gravity in a quantitative way.

MC realizations of $\phi_0$, $\phi$, $i$ and $e$ follow those described in \cite{chae2023}. They can be summarized as follows: (1) $\phi_0$ is drawn randomly from the range $(0,2\pi)$; (2) $\phi$ comes from the time along the orbit randomly drawn from $(0,T)$ where $T$ is the period; (3) $i$ is randomly drawn from $(0,\pi/2)$ with a probability density $p(i)=\sin(i)$; finally, (4) $e$ is drawn from the individual ranges provided by \cite{hwang2022}  {as shown in figure~8 of \cite{chae2023}}.  {Note particularly that the eccentricity distribution for each binary is specified by three values: the most likely value ($e_m$), a lower-bound value ($e_l$), and an upper-bound value ($e_u$). In an MC, eccentricity is drawn using a combination of two truncated Gaussian functions: $e_m$ is taken as the median and each side is assumed to follow a truncated Gaussian function with a ``$\sigma$'' of $e_u-e_m$ or $e_m-e_l$ with the total range bounded by the limit $0.001<e<0.999$. I also consider a power-law distribution $p(e;\alpha)=(1+\alpha)e^\alpha$ with a systematically varying}
  \begin{equation}
    \alpha =-5.123 + 4.612 x -1.098 x^2 + 0.08748 x^3
  \label{eq:alpha}
 \end{equation}
 {with $x\equiv \log_{10}(s/\text{au})$ based on figure~7 and table~1 of \cite{hwang2022}. Equation~(\ref{eq:alpha}) is valid for the range $2\la x\la 4.5$.} For further details, the reader is referred to \cite{chae2023}.

\section{Results} \label{sec:result}

\begin{figure*}
  \centering
  \includegraphics[width=0.8\linewidth]{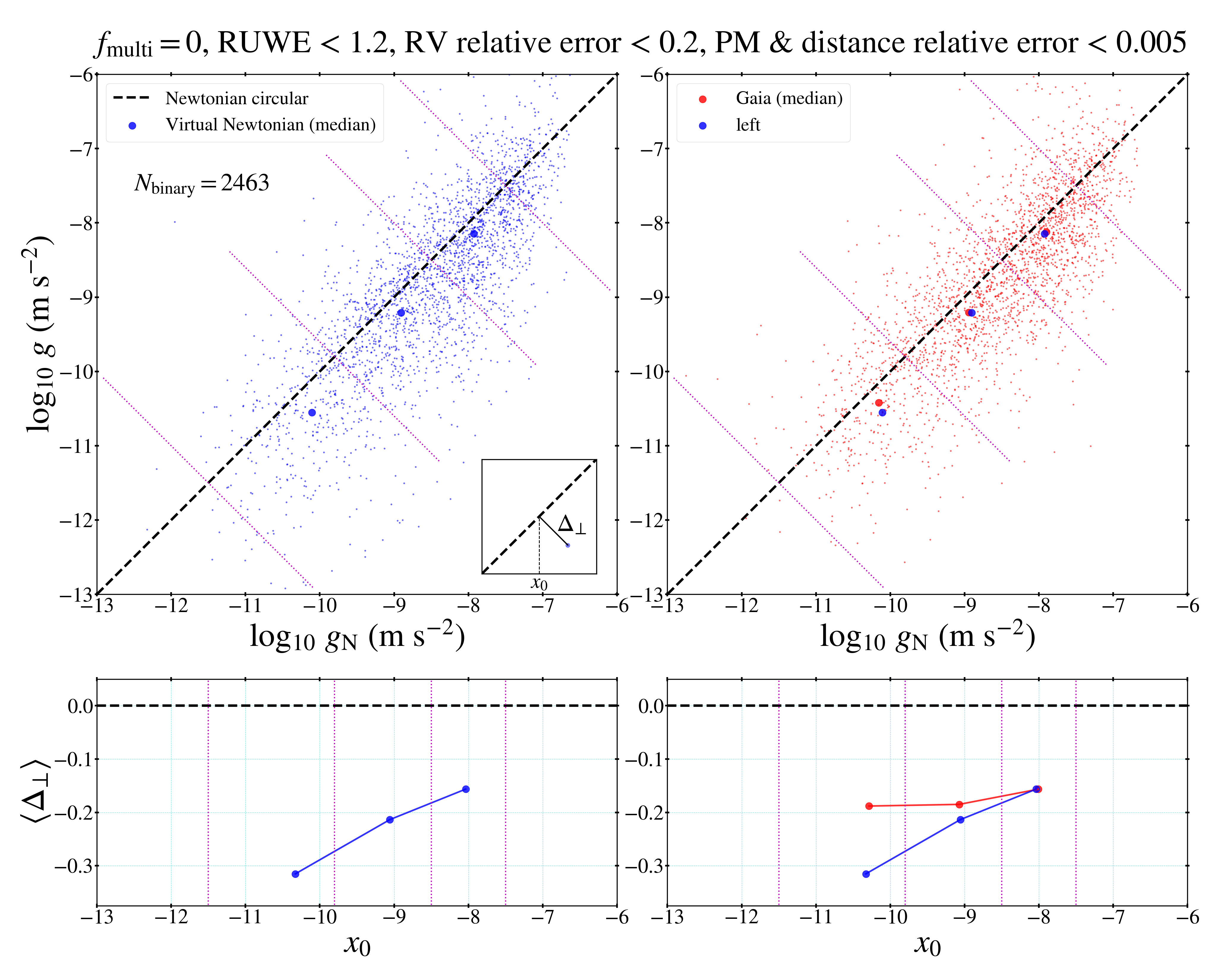}
    \vspace{-0.2truecm}
    \caption{\small 
    MC realized distributions of  {2,463} pure binaries in the acceleration plane defined in \cite{chae2023}. The quantity $g_{\rm{N}}\equiv GM_{\rm{tot}}/r^2$ is the Newtonian gravitational acceleration between the two stars and $g\equiv v^2/r$ is an empirical kinematic acceleration, where $r$ and $v$ are deprojected 3D separation and relative velocity. The left panel shows a Newton-predicted distribution while the right panel shows a distribution from Gaia measurements from \emph{one} MC realization. Big dots indicate the medians in the orthogonal bins indicated by magenta dotted lines. The orthogonal deviation $\Delta_\bot$ of a point from the diagonal line is indicated in the inset of the upper left panel. The bottom panels show the medians of $\Delta_\bot$ in the bins. In the right panels, the Gaia medians are compared with he Newtonian medians.
    } 
   \label{RAR_main}
\end{figure*} 

The sample of  {2,463} pure binaries defined in Section~\ref{sec:data} was obtained from a systematic investigation using the code developed in \cite{chae2023}  {and revised as described in Appendix~\ref{sec:correction}}. Various samples defined in \cite{chae2023} always had $f_{\rm{multi}}>0$ when binary motions were fitted to the Newtonian expectation in the high-acceleration regime $\ga 10^{-8}$m~s$^{-2}$. This means that generally defined samples always include undetected close companions as widely appreciated. Also, the values of $f_{\rm{multi}} \approx 0.2 - 0.5$ obtained in \cite{chae2023}  {with the revision of Appendix~\ref{sec:correction}} agree well with the results from various surveys \citep{raghavan2010,tokovinin2014,riddle2015,moe2017} indicating that the calibration procedure of \cite{chae2023} is reliable.

  In Section~\ref{sec:accel}, I show the results in the acceleration plane for the sample of pure binaries obtained with the code of \cite{chae2023}. The results clearly show that binary motions match the Newtonian expectation with $f_{\rm{multi}}=0$ in the high-acceleration regime. The results will then provide a new test of gravity in the low-acceleration regime. In Section~\ref{sec:velprof}, I present the main results of this work, i.e., the stacked velocity profiles compared with the MC predictions of Newtonian gravity.
  
\subsection{Results on the acceleration plane} \label{sec:accel}  

Figure~\ref{RAR_main} shows one MC result for the kinematic acceleration $g\equiv v^2/r$ defined in \cite{chae2023} against the Newtonian gravitational acceleration $g_{\rm{N}}\equiv G M_{\rm{tot}}/r^2$ for the main sample of  {2,463} pure binaries. Here stellar masses are estimated through the standard magnitude-mass ($M_G$-$M$) relation (the first choice in table~1 of \cite{chae2023}) that is based on the \cite{pecaut2013} $V$-band magnitude-mass relation. The Newtonian expectation of the $g_{\rm{N}}$-$g$ relation is compared with that for the Gaia data. In particular, the median orthogonal deviations $\langle\Delta_\bot\rangle$ in the acceleration bins are quantitatively compared, as shown in the bottom panels of Figure~\ref{RAR_main}.

\begin{figure*}
  \centering
  \includegraphics[width=0.8\linewidth]{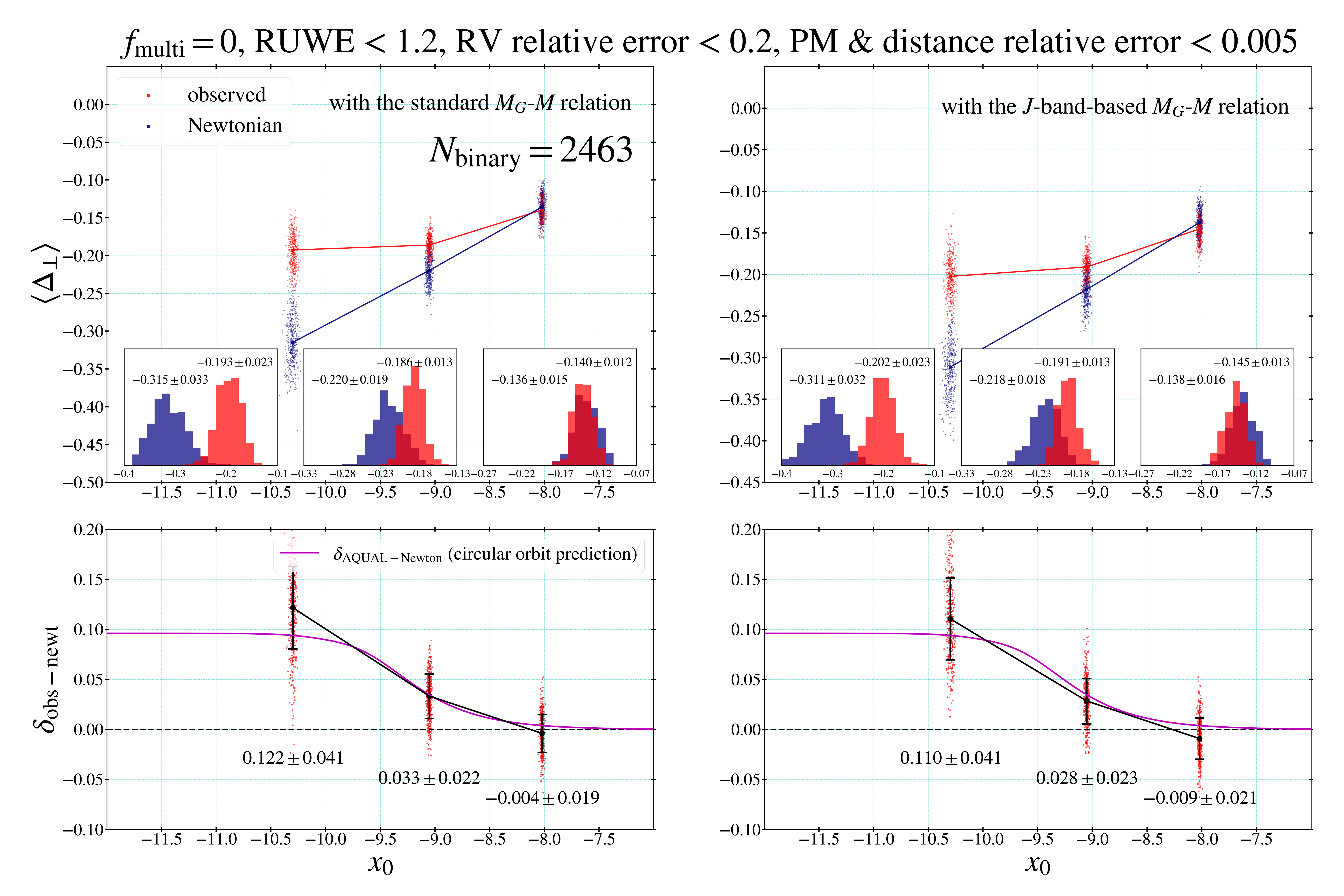}
    \vspace{-0.2truecm}
    \caption{\small 
    The upper left panel shows distributions of median orthogonal deviations $\langle\Delta_\bot\rangle$ (defined in Figure~\ref{RAR_main}) from 400 MC results with the standard $M_G$-$M$ relation for the main sample of  {2,463} pure binaries. The upper right panel is with the $J$-band-based $M_G$-$M$ relation. The bottom panels show distributions of the difference $\delta_{\rm{obs-newt}}\equiv \langle\Delta_\bot\rangle_{\rm{obs}} - \langle\Delta_\bot\rangle_{\rm{N}}$. In the lowest acceleration bin at $x_0\approx -8.0$,  $\delta_{\rm{obs-newt}} = 0$ is naturally satisfied without any adjustment of $f_{\rm{multi}}$. The magenta curve in the bottom panels represents the AQUAL prediction for circular orbits with the Milky Way external field (see  {Figure~\ref{EFE_MW}}).
    } 
   \label{delg_main}
\end{figure*} 

Another MC gives different $\langle\Delta_\bot\rangle$ values, and probability distributions of $\langle\Delta_\bot\rangle$ can be derived from a number of MC results as demonstrated in \cite{chae2023}. Figure~\ref{delg_main} shows the distributions from 400 MC results. Those shown in the left column are from the MC results with the standard $M_G$-$M$ relation while those in the right are with the $J$-band based $M_G$-$M$ relation (the second choice in table~1 of \cite{chae2023}). It is clearly seen that the Gaia result naturally matches the Newtonian expectation in the highest acceleration bin at $x_0\approx -8$ with $f_{\rm{multi}}=0$ (i.e.\ without any calibration) whichever $M_G$-$M$ relation is used. The parameter $\delta_{\rm{obs-newt}} \equiv \langle\Delta_\bot\rangle_{\rm{observed}} - \langle\Delta_\bot\rangle_{\rm{Newton}}$ defined in paper~I has values of $\delta_{\rm{obs-newt}} = -0.004\pm 0.019$ and $-0.009\pm 0.021$ at $x_0\approx -8$, well consistent with zero. This agreement is remarkable considering that there are no free parameters.

\begin{figure*}
  \centering
  \includegraphics[width=0.8\linewidth]{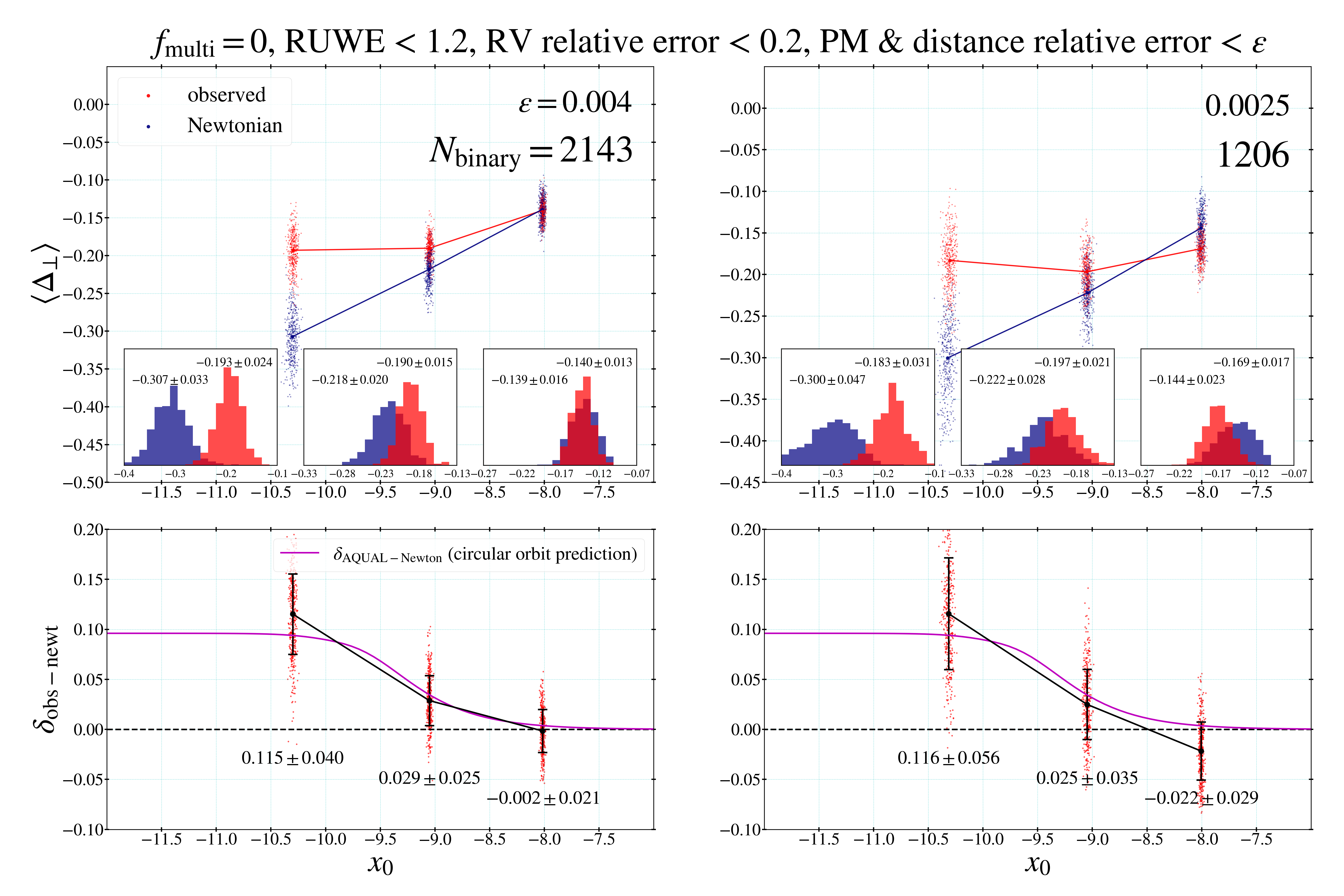}
    \vspace{-0.2truecm}
    \caption{\small 
    The same as the left panel of Figure~\ref{delg_main} but for subsamples with stricter astrometric requirements than the main sample. The results have larger statistical uncertainties and are consistent with those for the main sample. 
    } 
   \label{delg_epsilon}
\end{figure*}

To check that the above agreement with the Newtonian expectation at $x_0\approx -8$ in the main sample is true rather than a coincidence, I consider subsamples selected with more stringent astrometric requirements (along with the same requirement on radial velocities). When relative errors of PMs and parallaxes are required to be $<\varepsilon$, I consider $\varepsilon=0.004$ and $0.0025$. The MC results on the acceleration plane for these subsamples can be found in Figure~\ref{delg_epsilon}. These results agree well with $\delta_{\rm{obs-newt}} = 0$ at $x_0\approx -8$ though with larger statistical errors due to smaller sample sizes. 

\begin{figure}
  \centering
  \includegraphics[width=0.9\linewidth]{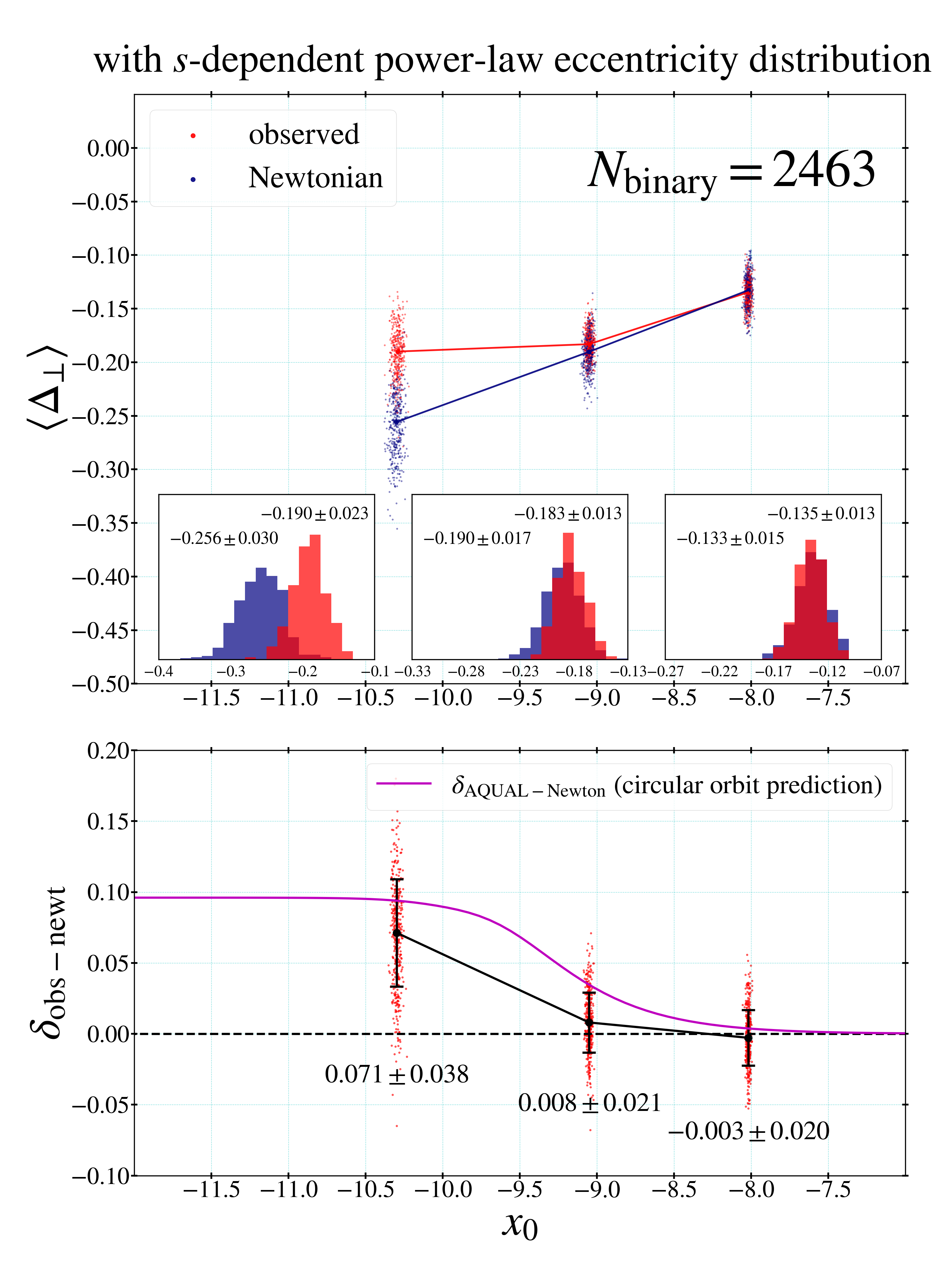}
    \vspace{-0.2truecm}
    \caption{\small 
     {The same as the left panel of Figure~\ref{delg_main} but with statistical eccentricities drawn from a power-law distribution with a varying index given by Equation~(\ref{eq:alpha}).}
    } 
   \label{delg_stateccen}
\end{figure}

 {The above results are based on individual eccentricities estimated by \cite{hwang2022} and thus represent currently most likely results. Figure~\ref{delg_stateccen} shows a result based on eccentricities drawn statistically from a power-law distribution with a varying index given by Equation~(\ref{eq:alpha}). Because binary-specific eccentricities are replaced by statistical eccentricities, the deviation is somewhat diluted as already noted in \cite{chae2023}. However, the result still favors the AQUAL model over Newton.} 

\begin{figure*}
  \centering
  \includegraphics[width=0.8\linewidth]{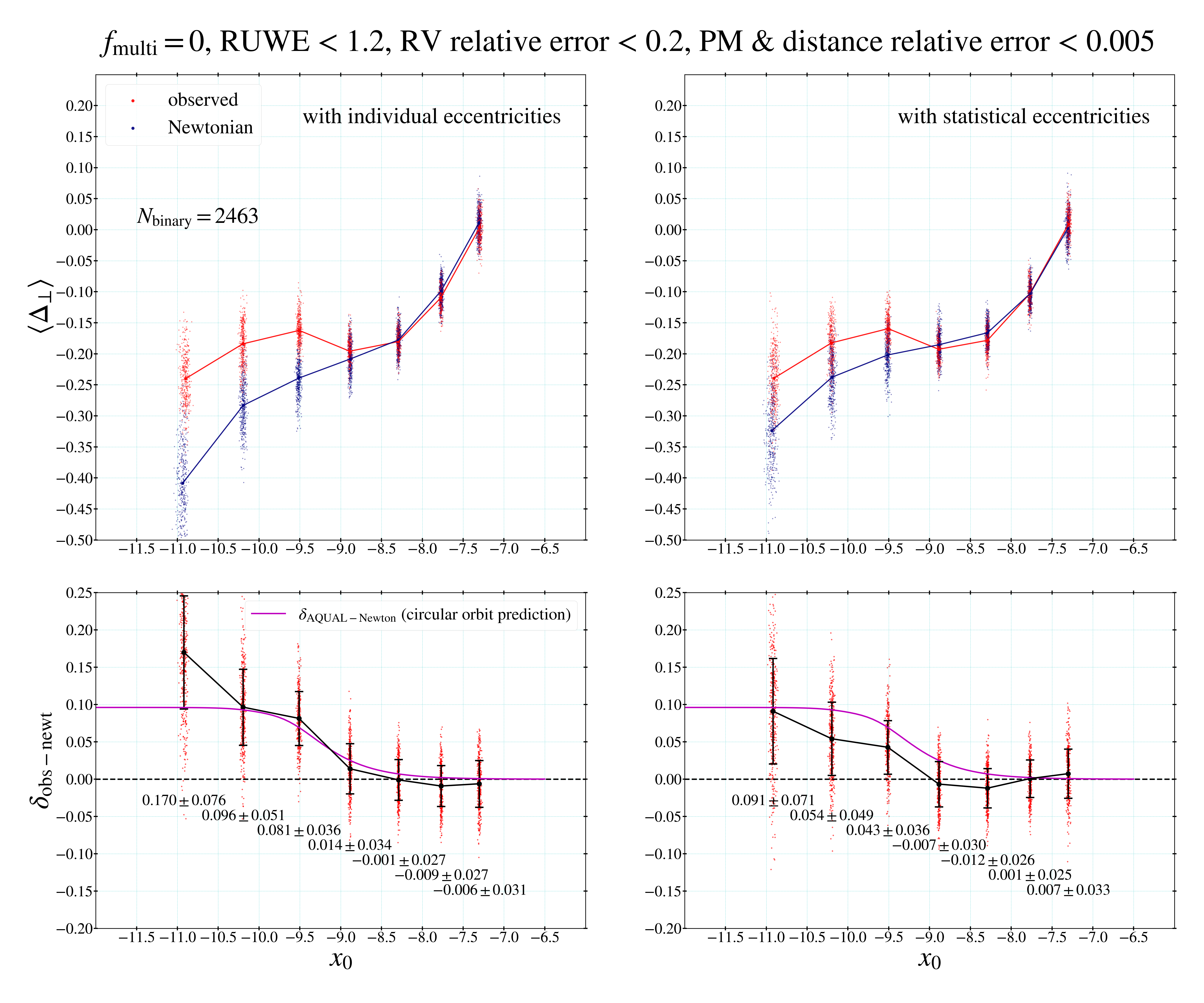}
    \vspace{-0.2truecm}
    \caption{\small 
       {Similar to Figure~\ref{delg_main} but with 7 bins. The left panel is with the standard inputs including individual eccentricities while the right panel is with statistical eccentricities replacing individual eccentricities.}
      } 
   \label{delg_7bins}
\end{figure*}

 {So far I have considered three bins so that each bin has a maximal number of MC points for bins of significantly different accelerations. Now I consider finer bins to test any dependence of the results with binning. Figure~\ref{delg_7bins} shows the results. The result with the standard inputs including individual eccentricities follows the AQUAL curve remarkably well over the entire bins. }

If the astrometric requirements or the requirement on radial velocities are relaxed progressively from PM and parallax relative errors $< 0.005$ or RV relative errors $<0.2$, one can check that $\delta_{\rm{obs-newt}}$ at $x_0 = -8$ with $f_{\rm{multi}}=0$ increases progressively from zero. Note that the astrometric and RV requirements can be less stringent than the presently chosen requirements depending on the tolerance about the value of $\delta_{\rm{obs-newt}}$ at $x_0 = -8$ with $f_{\rm{multi}}=0$. For example, one could allow $\delta_{\rm{obs-newt}}$ to be consistent with zero only within the MC estimated $1\sigma$. Here I am very conservative and require $\delta_{\rm{obs-newt}}$ to be consistent with zero within a small fraction of the MC estimated $1\sigma$.

The above results verify that the main sample with the presently chosen astrometric and RV requirements is \emph{statistically} free of hierarchical systems. Of course, there still can be a few individual systems that have minor undetected close companions. However, even if they are present, they are statistically negligible. The above results also reassure that the whole procedure, the Gaia data, and the empirical $M_G$-$M$ relations are all reliable. 

Now the derived values of $\delta_{\rm{obs-newt}}$ at lower accelerations are not consistent with zero indicating that Newtonian gravity breaks down in the low-acceleration regime. Because the same astrometric and RV requirements are imposed on all binaries regardless of the separation $s$, it is unreasonbale to imagine that only more widely ($s\ga 2$~kau) separated binaries preferentially have large amounts of undetected companions while the less widely ($s\la 1$~kau) separated binaries have none. Thus, these results provide robust evidence for two aspects of gravity: (1) Newtonian gravity holds for acceleration $\ga 10^{-8}$~m~s$^{-2}$, (2) Newtonian gravity (and thus general relativity) breaks down in the low acceleration regime $\la 10^{-9}$~m~s$^{-2}$.

While the evidence from statistically pure binaries is robust, the statistical significance is much weaker than in \cite{chae2023} due to the much smaller sample size. From the results with the standard $M_G$-$M$ relation, $\delta_{\rm{obs-newt}}=0.122\pm 0.041$ at $x_0\approx -10.3$ with a significance of $3.0\sigma$. At $x_0\approx -9.1$, $\delta_{\rm{obs-newt}}=0.033\pm 0.022$ with a significance of $1.5\sigma$. Taken together these results rule out Newtonian gravity with a $>3\sigma$ confidence. At $x_0\approx -10.3$, the acceleration boost factor is
 \begin{equation}
  \gamma_g \equiv \frac{g_{\rm{obs}}}{g_{\rm{pred}}} = 10^{\sqrt{2}\delta_{\rm{obs-newt}}} = 1.49_{-0.19}^{+0.21},
  \label{eq:gamma_g}
 \end{equation}
where $g_{\rm{obs}}=v^2 /r$ is the MC realized kinematic acceleration from the Gaia data, and $g_{\rm{pred}}$ is the corresponding Newtonian prediction. This value is in excellent agreement with the values obtained in \cite{chae2023} for general samples with $f_{\rm{multi}}$ self-calibrated.

\subsection{Profiles of stacked velocities} \label{sec:velprof}  
  
Here I present profiles of stacked velocities for the pure binary sample and compare them with the Newtonian MC predictions as described in Section~\ref{sec:method}. I pay most attention to sky-projected velocity ($v_p$) profiles because radial velocities ($v_r$) as precise as $v_p$ are quite rare. However, I will also consider, for the first time,  profiles of physical velocities $v$ (Equation~(\ref{eq:v_ob})) for a few dozens of binaries with exceptionally precise $v_r$ measurements.

\begin{figure*}
  \centering
  \includegraphics[width=0.7\linewidth]{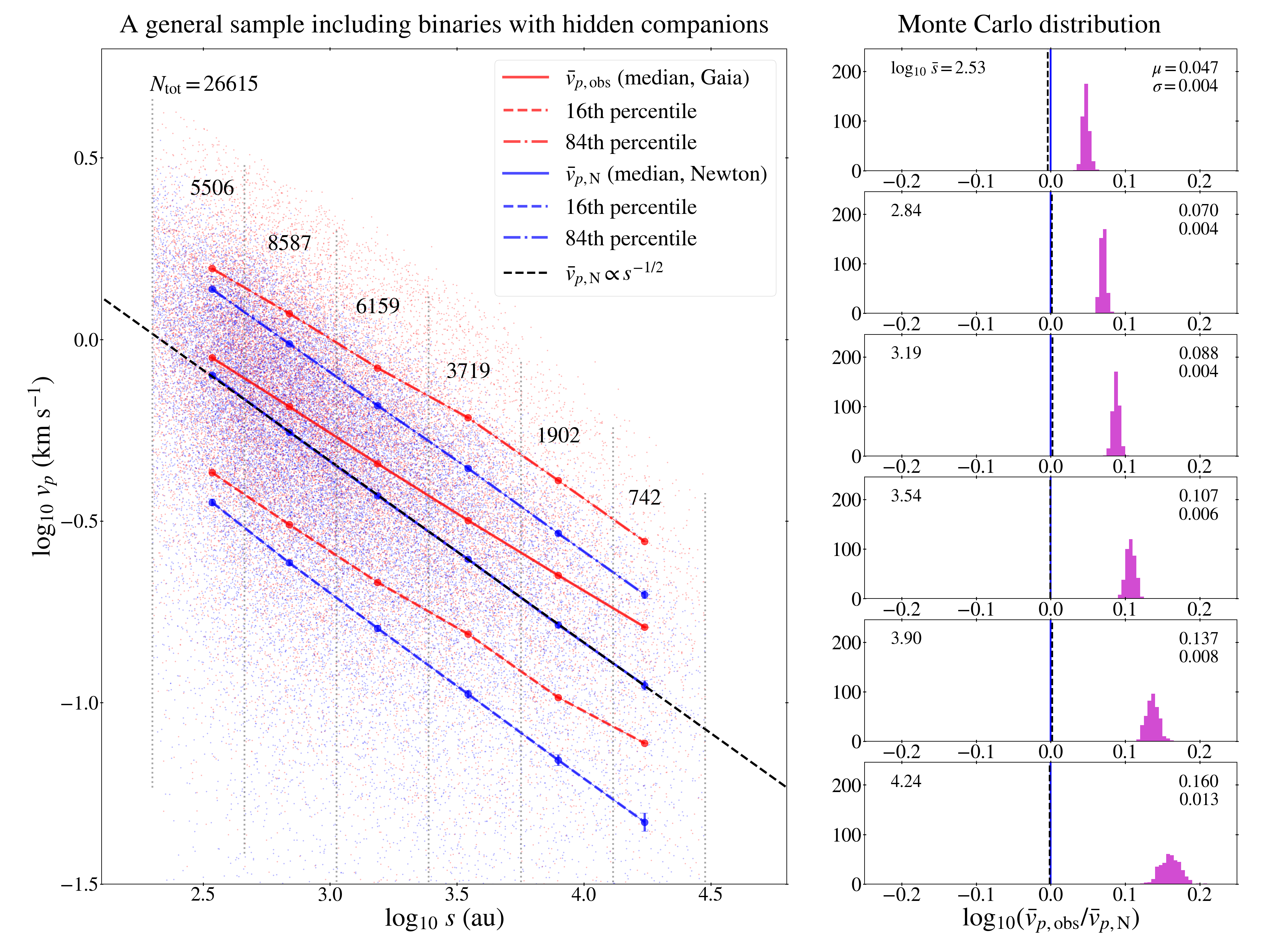}
    \vspace{-0.2truecm}
    \caption{\small 
     {The left panel shows sky-projected velocities with respect to sky-projected separation $s$ for the general sample of 26,615 binaries with the standard $M_G$-$M$ relation. Small red dots are the observed velocities (Equation~(\ref{eq:vp_ob})) while blue ones are Newtonian velocities (Equation~(\ref{eq:vpN_s})) in \emph{one} MC realization. Big dots indicate median, 16th, and 84th percentile velocities in the bins of $s$ defined in Figure~\ref{massdist}. The Newtonian median velocities $\bar{v}_{p,{\rm{N}}}$ follow the Keplerian scaling of $\bar{v}_{p,{\rm{N}}}\propto s^{-1/2}$. The observed median velocities deviate from the Newtonian predictions in the entire range and the scaling as a slope clearly different from $-1/2$.  The right panels show the probability distributions of $\log_{10}(\bar{v}_{p,{\rm{obs}}}/\bar{v}_{p,{\rm{N}}})$ derived from 400 MC realizations.}  
    } 
   \label{vp_impure}
\end{figure*} 

\subsubsection{Testing the general sample including impure binaries} \label{sec:impure}

 {Before presenting results on pure binaries I test the general sample of 26,615 binaries defined in \cite{chae2023} that includes ``impure'' binaries hosting unresolved hidden close companions. Figure~\ref{vp_impure} compares the observed sky-projected velocities with the Newton-predicted values without taking into account hidden close companions. As expected, the observed velocities are higher than the Newton-predicted values regardless of the separation between the two stars. This is a clear indication that the observed velocities are largely affected by additional masses from hidden companions.}

 {However, Figure~\ref{vp_impure} also reveals that the observed binned medians do not follow the Keplerian scaling of $\propto s^{-0.5}$. The measured scaling has a slope of $-0.437\pm 0.003$ exhibiting a $21\sigma$ deviation from $-0.5$. This indicates that $f_{\rm{multi}}$ must increase sharply with $s$ to be consistent with the Keplerian scaling if gravitational anomaly or other factors are not permitted. I note that any bias arising from distances cannot explain this slope because a subsample with a narrow distance range shows a similar slope. Thus, we are in a situation where ``general'' binaries at the same distances selected with the same criteria show stong differences in kinematics depending only on $s$. This motivates investigations of ``pure'' binaries in the following sections. }

\begin{figure*}
  \centering
  \includegraphics[width=0.7\linewidth]{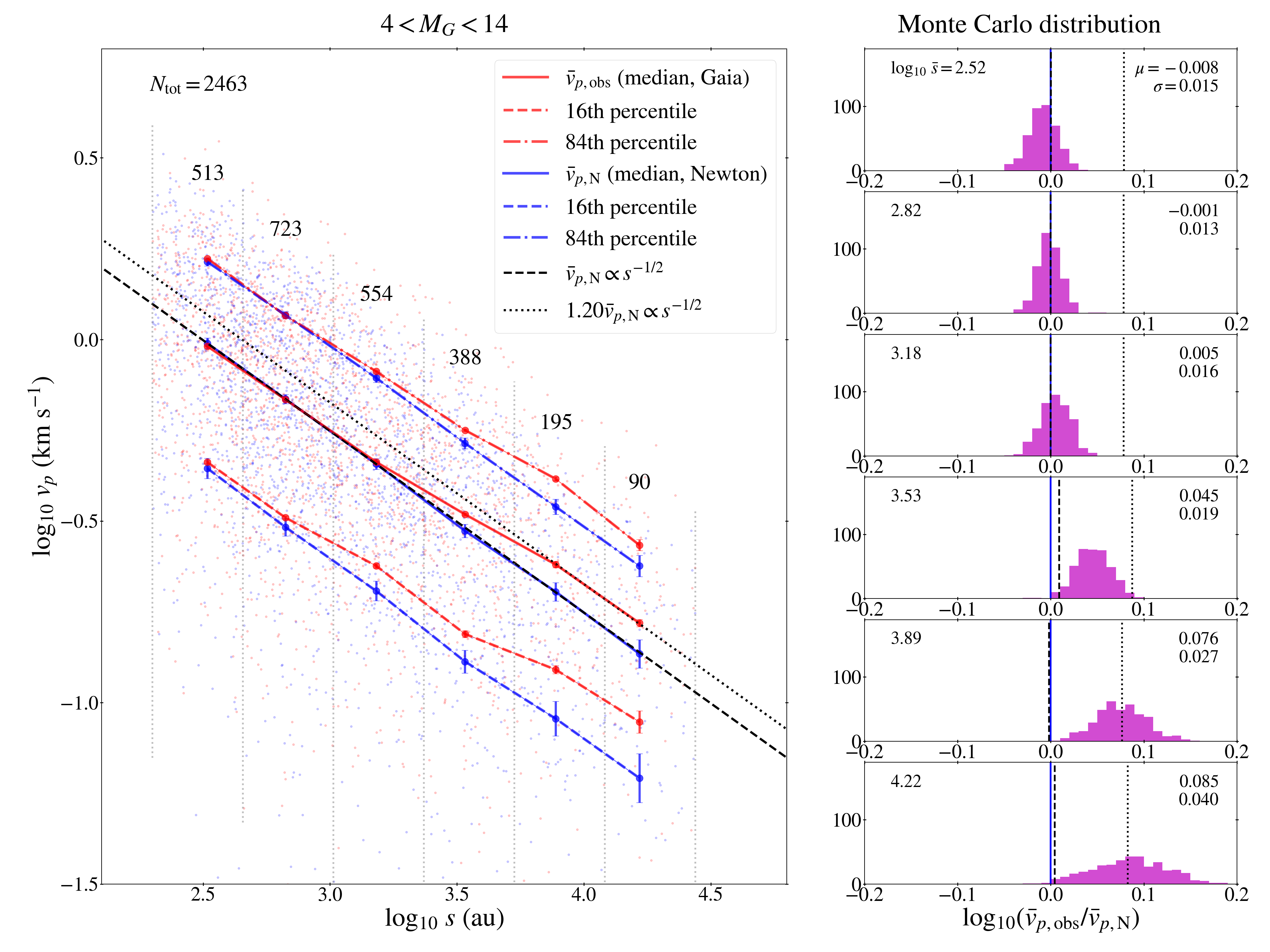}
    \vspace{-0.2truecm}
    \caption{\small 
    The left panel shows sky-projected velocities with respect to sky-projected separation $s$ for the main sample of  {2,463} pure binaries with the standard $M_G$-$M$ relation. Small red dots are the observed velocities (Equation~(\ref{eq:vp_ob})) while blue ones are Newtonian velocities (Equation~(\ref{eq:vpN_s})) in \emph{one} MC realization. Big dots indicate  {the median, the 16th percentile, and the 84th percentile velocities in the bins.} The error bars of the big  {blue} dots are estimated from 400 MC realizations. The Newtonian median velocities $\bar{v}_{p,{\rm{N}}}$ follow the Keplerian scaling of $\bar{v}_{p,{\rm{N}}}\propto s^{-1/2}$.  {The observed median velocities $\bar{v}_{p,{\rm{obs}}}$ in the three lowest-$s$ bins naturally match the Newtonian predictions.} However, the observed median velocities deviate from the Newtonian predictions in the larger-$s$ bins. The dotted line indicates the boosted velocity in the two largest-$s$ bins. The right panels show the probability distributions of $\log_{10}(\bar{v}_{p,{\rm{obs}}}/\bar{v}_{p,{\rm{N}}})$ derived from the MC realizations.  
    } 
   \label{vp_main}
\end{figure*} 

\subsubsection{Main results} \label{sec:main} 

Figure~\ref{vp_main} shows the profile of stacked $v_p$ values for all  {2,463} pure binaries in the main sample and compares it with the Newtonian prediction with the standard $M_G$-$M$ relation. Small red dots represent individual observed values  {while the big red dots represent the median, the 16th percentile, and the 84th percentile in the bins of $s$.} Small blue dots represent an output from one Newtonian MC run. The mean values of  {the median, the 16th percentile, and the 84th percentile} in the bins of $s$ are obtained from 400 MC runs. The big blue dots and errorbars shown in the left panel of Figure~\ref{vp_main} represent the distributions from MC runs. Note that the Newtonian MC predicted velocities can have tangible uncertainties when numbers are small as in the relatively larger-$s$ bins. Histograms in the right panels of Figure~\ref{vp_main} show the distributions of $\log_{10}(\bar{v}_{p,\rm{obs}}/\bar{v}_{p,\rm{N}})$ in the  {6} bins. Note that the widths of the distributions are entirely determined by the distributions of $\bar{v}_{p,\rm{N}}$. 

Figure~\ref{vp_main} shows that in the  {three smallest-$s$ bins the observed median velocities agree well with the Newton-predicted values with $\log_{10}(\bar{v}_{p,\rm{obs}}/\bar{v}_{p,\rm{N}})=-0.008\pm 0.015$, $-0.001\pm 0.013$, and $0.005\pm 0.016$. Moreover, the 16th and 84th percentiles of the observed velocities agree well with the Newton-predicted values for the first two bins indicating that the observed distributions are fully consistent with the Newton-predicted distributions.} Thus, for binary systems with $0.5\la M_{\rm{tot}}/{\rm{M}}_\odot \la 2$ Newtonian dynamics holds for $s\la 2$~kau  {as expected from Figures~\ref{EFE_MW} and \ref{transition}}. This result is consistent with the results on the acceleration plane presented in Section~\ref{sec:accel}.

The Newton-predicted median velocities follow the Keplerian profile $\propto s^{-1/2}$ as expected because the median mass varies little with $s$ in the sample as shown in Figure~\ref{massdist}. However, the observed median velocities do not follow the Newtonian profile  {for the probed entire $s$ range}. There is a jump in the $\bar{v}_{p,\rm{obs}}$ profile around $s\approx 2$~kau.   {The deviation from the Newtonian predictions in the 4th bin is $\log_{10}(\bar{v}_{p,\rm{obs}}/\bar{v}_{p,\rm{N}})=0.045\pm 0.019$ at $\log_{10} (\bar{s}/\text{au})=3.53$. However, in the last two bins the deviations are $0.076\pm 0.027$ and $0.085\pm 0.040$ at $\log_{10} (\bar{s}/\text{au})=3.89$ and $4.22$.} These deviations together represent  {a $\approx 5.0\sigma$ detection} of gravitational anomaly.  {Hereafter the statistical significance is estimated as follows. For the last three bins that deviate from the solid blue line, the probabilities $p(x<0)$ (where $x\equiv \log_{10}(\bar{v}_{p,\rm{obs}}/\bar{v}_{p,\rm{N}})$) are calculated based on the estimated $\mu$ and $\sigma$ values and the product of the three probabilities is obtained. Then, the product value is used to estimate a Gaussian equivalent significance.}

In the two largest-$s$ bins of Figure~\ref{vp_main} the boost factor for projected velocities is
 \begin{equation}
  \gamma_{v_p} \equiv \frac{\bar{v}_{p,\rm{obs}}}{\bar{v}_{p,\rm{N}}} = 1.20\pm 0.06 \text{ (with individual {\it{e}})}.
  \label{eq:gamma_vp}
 \end{equation}
 Because the kinematic acceleration was defined to be $v^2/r$, $\gamma_{v_p}$ is expected to match $\sqrt{\gamma_g}$ if projection effects are averaged out in a statistical sample. This is indeed the case from Equations~(\ref{eq:gamma_g}) and (\ref{eq:gamma_vp}). 

 A qualitatively important aspect of the stacked velocity profile shown in Figure~\ref{vp_main} is that the last two bins follow the Keplerian scaling $\propto s^{-1/2}$ indicating that they are pseudo-Newtonian with a boosted gravity as exactly predicted by MOND gravity under an EFE as shown in Figure~\ref{EFE_MW}. Thus, Figure~\ref{vp_main} agrees well with the unique trait of MOND gravity. 

\begin{figure*}
  \centering
  \includegraphics[width=0.7\linewidth]{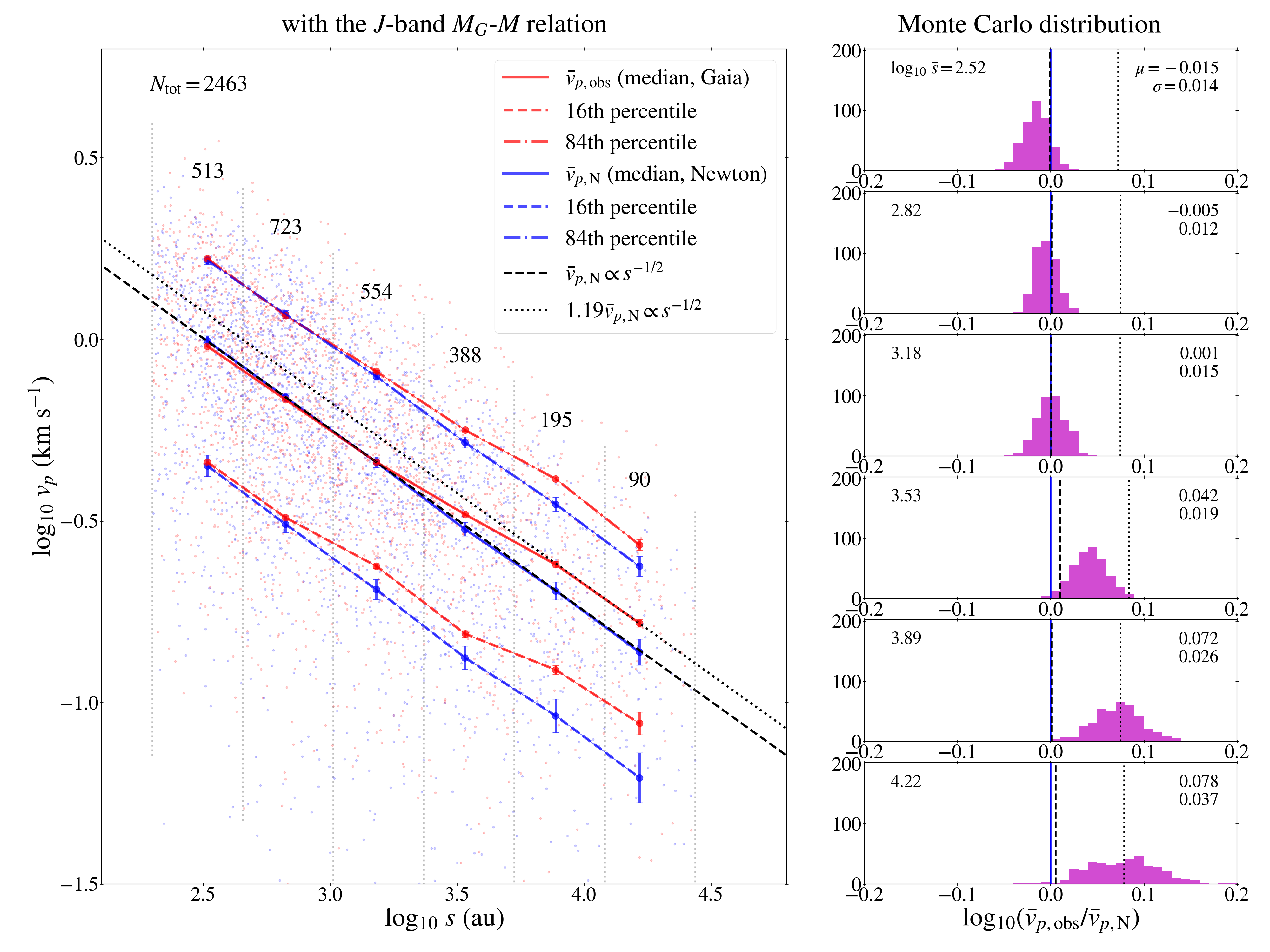}
    \vspace{-0.2truecm}
    \caption{\small 
    The same as Figure~\ref{vp_main} but with the $J$-band-based $M_G$-$M$ relation.
    } 
   \label{vp_main_j}
\end{figure*}

Figure~\ref{vp_main_j} shows the result with the $J$-band-based $M_G$-$M$ relation. This result is consistent with the result with the standard $M_G$-$M$ relation  {indicating that the results are robust within the currently available $M_G$-$M$ relations. The significance of the gravitational anomaly from the last three bins is $4.9\sigma$. However, below I will further probe the effects of systematically varying the observed Gaia magnitudes.}  
 
\begin{figure*}
  \centering
  \includegraphics[width=0.7\linewidth]{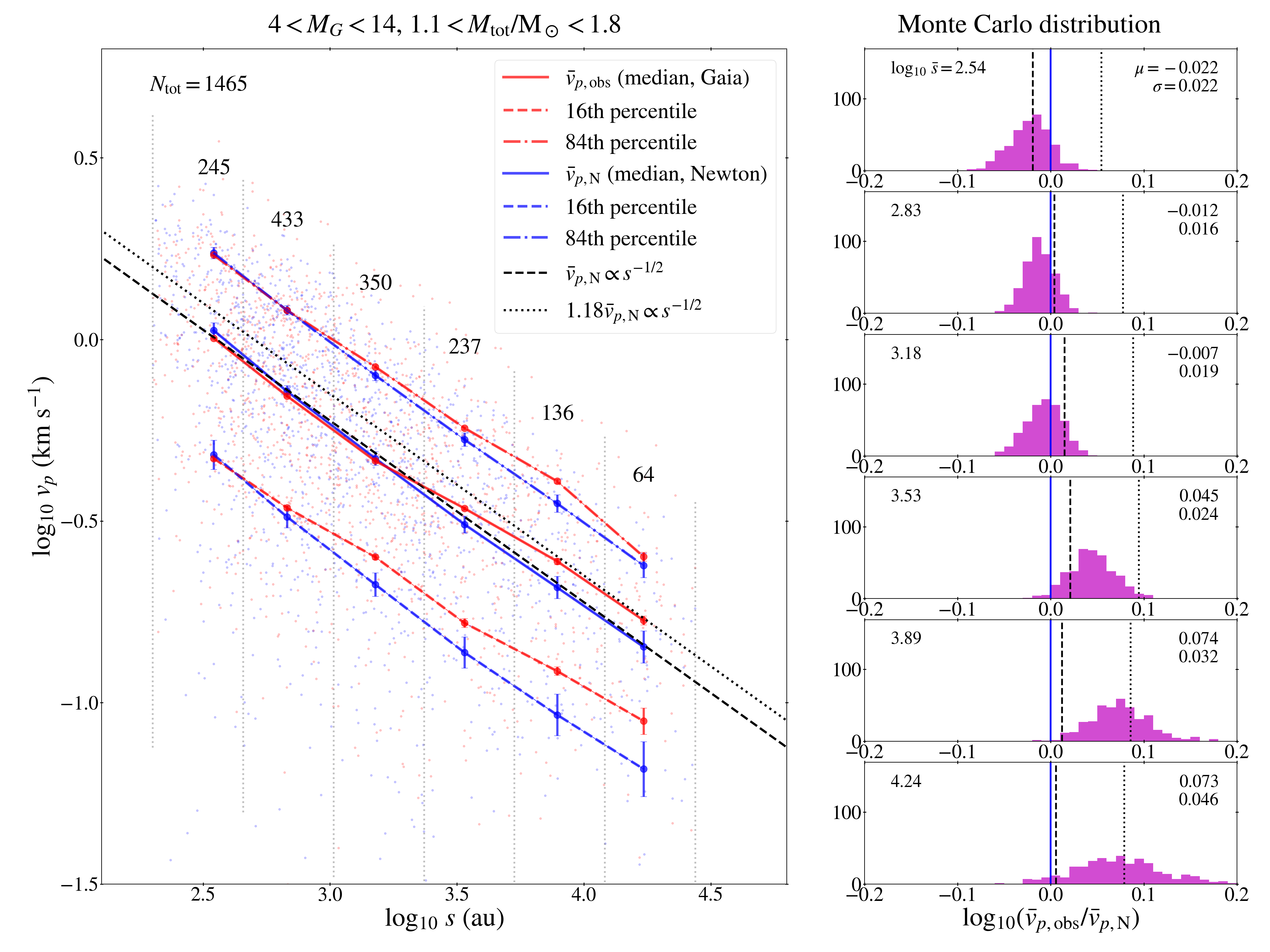}
    \vspace{-0.2truecm}
    \caption{\small 
    The same as Figure~\ref{vp_main} but for the subsample with the limited mass range $1.1<M_{\rm{tot}}/{\rm{M}}_\odot<1.8$.
    } 
   \label{vp_main_massltd}
\end{figure*}

\begin{figure*}
  \centering
  \includegraphics[width=0.7\linewidth]{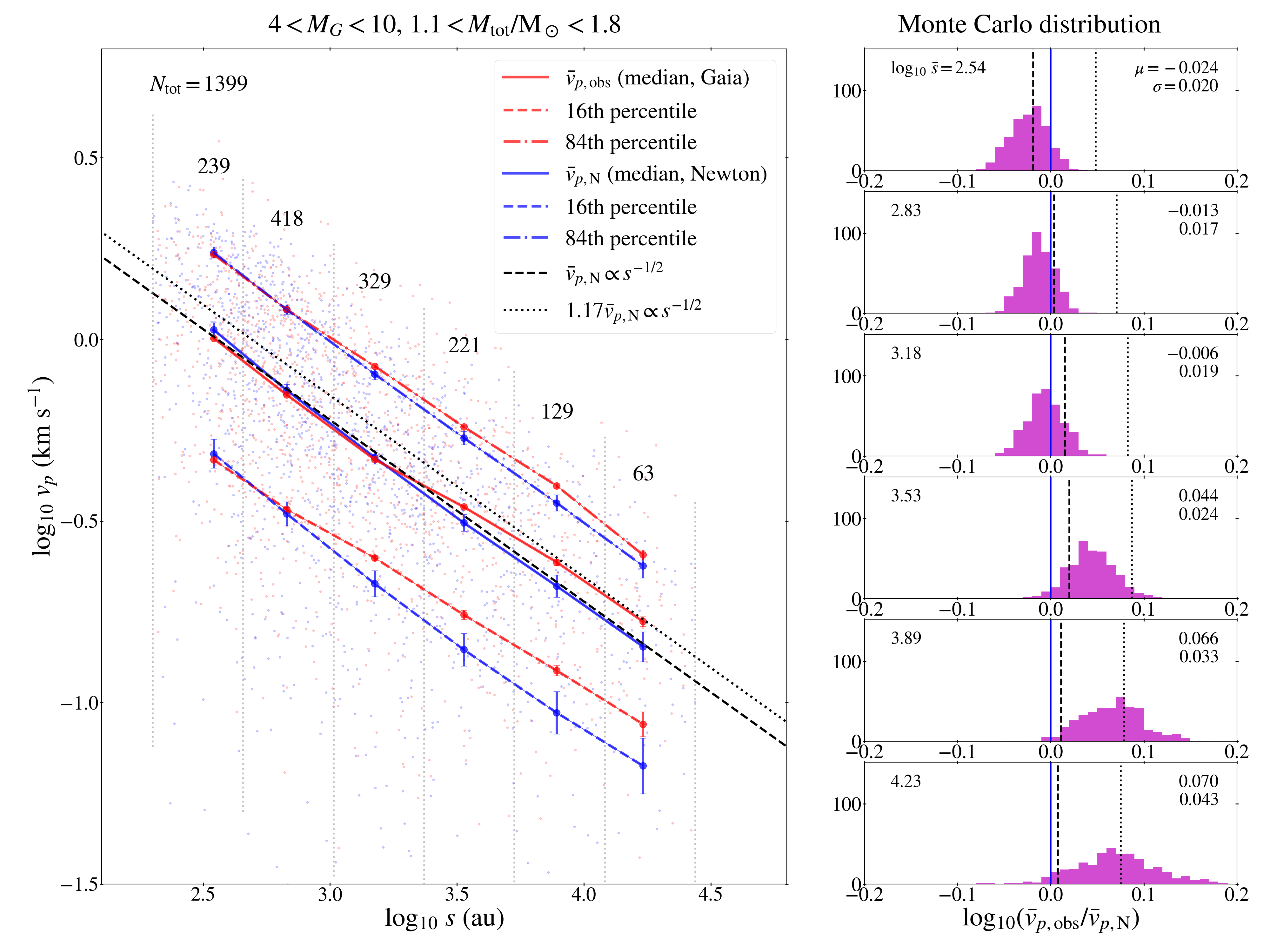}
    \vspace{-0.2truecm}
    \caption{\small 
    The same as Figure~\ref{vp_main} but for the subsample with the narrower magnitude range $4<M_G<10$ and the limited mass range $1.1<M_{\rm{tot}}/{\rm{M}}_\odot<1.8$.
    } 
   \label{vp_narrow_massltd}
\end{figure*}

The above results are for a mass range $0.5\la M_{\rm{tot}}/{\rm{M}}_\odot\la 2.5$. Now I consider subsamples with a limited mass range $1.1<M_{\rm{tot}}/{\rm{M}}_\odot<1.8$. Figure~\ref{vp_main_massltd} shows the results for the subsample with $1.1<M_{\rm{tot}}/{\rm{M}}_\odot<1.8$ and the clean magnitude range $4<M_G<14$. Figure~\ref{vp_narrow_massltd} shows the results for the subsample with $1.1<M_{\rm{tot}}/{\rm{M}}_\odot<1.8$ and a narrower magnitude range $4<M_G<10$.  {These results agree well with the results with the main sample. The statistical significance of the deviations is $\approx 4.0\sigma$ for both results.}

\begin{figure*}
  \centering
  \includegraphics[width=0.7\linewidth]{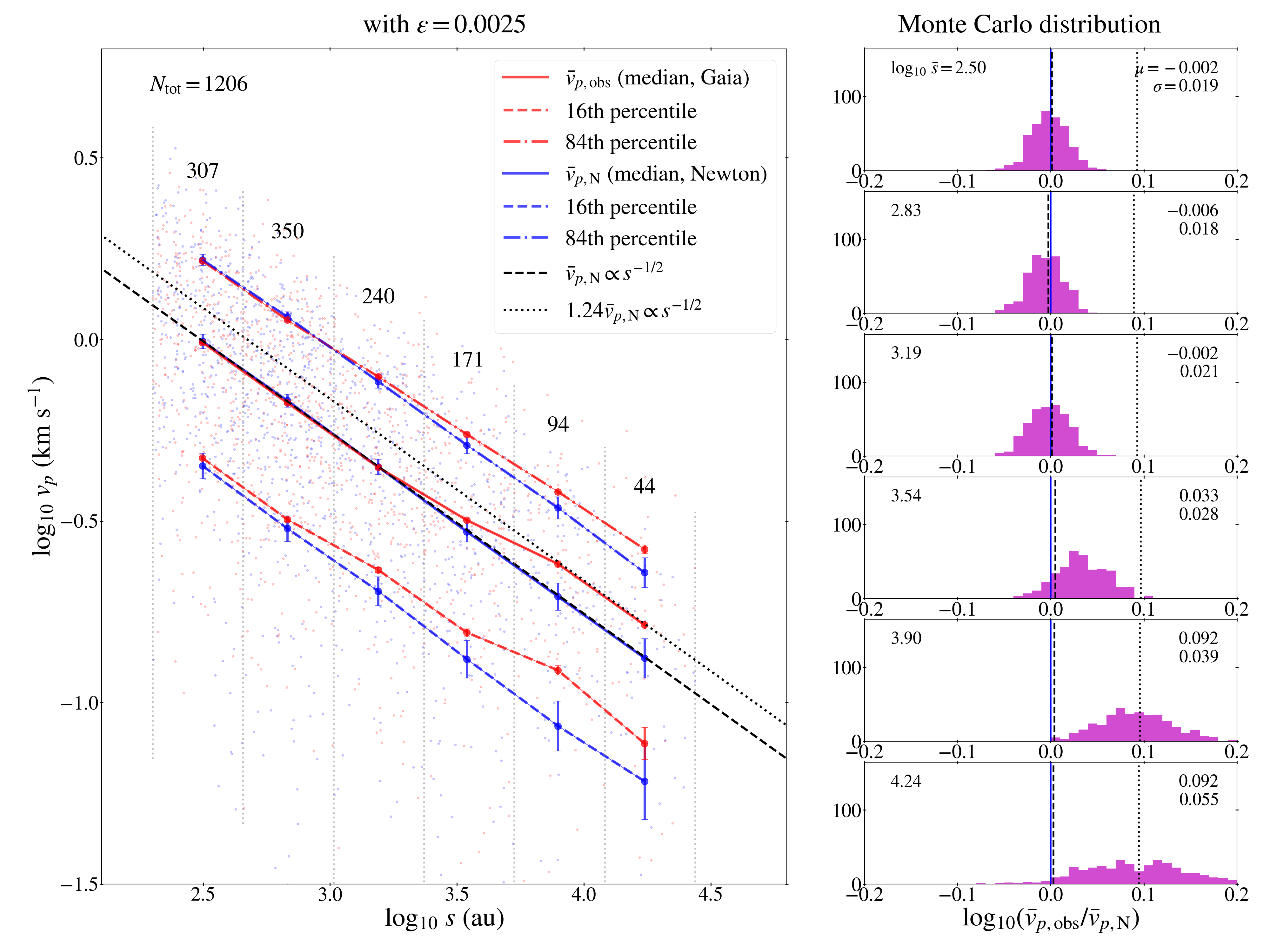}
    \vspace{-0.2truecm}
    \caption{\small 
     The same as Figure~\ref{vp_main} but for the subsample  {with a more stringent astrometric requirement of $\varepsilon=0.0025$.}
    } 
   \label{vp_eps0_0025}
\end{figure*}

It is interesting to consider subsamples obtained in the limiting cases of extreme precision of PMs and distances.  {Figure~\ref{vp_eps0_0025} shows the result for $1,206$ pure binaries with  {$\varepsilon = 0.0025$, i.e. twice better precision.} The result is well consistent with the results for the main sample. The statistical significance of the deviations in the last three bins taken together is $3.9\sigma$.}

 {Finally, I consider statistical eccentricities based on the power-law distribution with the slope systematically varying with $s$ as given by Equation~(\ref{eq:alpha}), rather than individual ranges of eccentricities reported by \cite{hwang2022}. The overall trend of the stacked velocity profile agrees well with the results with individual eccentricities. The statistical significance of the deviations in the last 3 bins taken together is $4.1\sigma$. The boost factor estimated based on the two largest-$s$ bins is slightly lower than the value given in Equation~(\ref{eq:gamma_vp}): }
 \begin{equation}
  \gamma_{v_p} \equiv \frac{\bar{v}_{p,\rm{obs}}}{\bar{v}_{p,\rm{N}}} = 1.15\pm 0.06 \text{ (with statistical {\it{e}})}.
  \label{eq:gamma_vp_state}
 \end{equation}

  {The main key results are summarized in Table~\ref{tab:main_result}. The projected velocity boost factor is in the range $1.15\le \gamma_{v_p} \le 1.24$. The statistical significance of the gravitational anomaly is in the range $(3.9\sigma,5.0\sigma)$. }
 
\begin{figure*}
  \centering
  \includegraphics[width=0.7\linewidth]{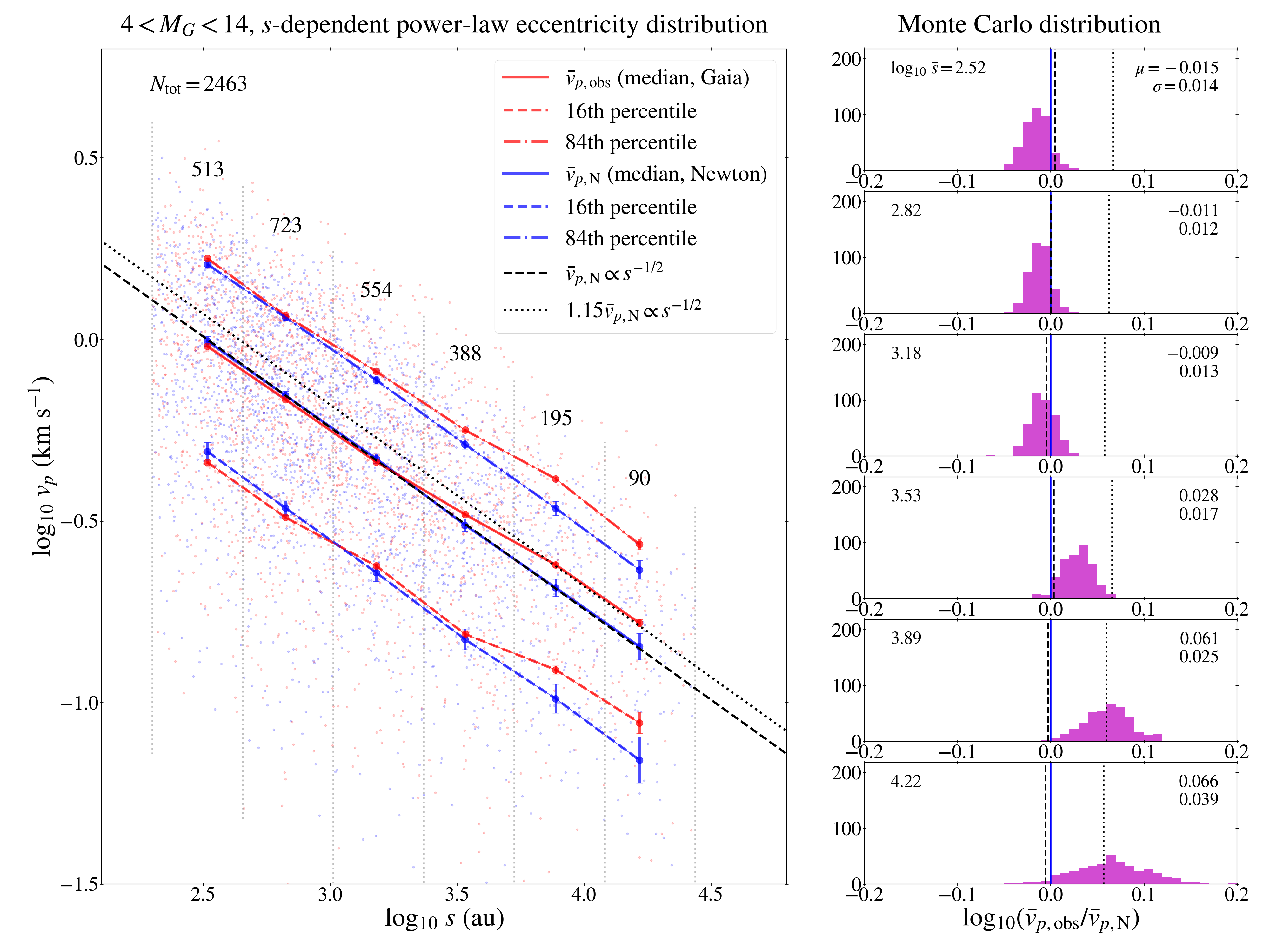}
    \vspace{-0.2truecm}
    \caption{\small 
     The same as Figure~\ref{vp_main} but with statistical eccentricities from an $s$-dependent power-law distribution.
    } 
   \label{vp_systematice}
\end{figure*} 

\begin{table*}
  \caption{ {Main results of gravitational anomaly from pure binaries}}\label{tab:main_result}
\begin{center}
  \begin{tabular}{cccccc}
  \hline
 sample   & $N_{\rm{binary}}$ &  precision cut & eccentricity & $\gamma_{v_p}$ ($s\ga 5$~kau)  & statistical significance ($s\ga 2$~kau)  \\
 \hline
 $4<M_G<14$, no limit on $M_{\rm{tot}}$ & 2463 &  $\varepsilon=0.005$ & individual  & $1.20\pm 0.06$ & $p(x<0)=3.1\times 10^{-7}$ ($5.0\sigma$)  \\
$4<M_G<14$, $1.1<M_{\rm{tot}}/{\rm{M}}_\odot<1.8$ & 1465 &  $\varepsilon=0.005$ & individual  & $1.18\pm 0.07$ & $p(x<0)=1.7\times 10^{-5}$ ($4.1\sigma$)  \\
$4<M_G<10$, $1.1<M_{\rm{tot}}/{\rm{M}}_\odot<1.8$ & 1399 &  $\varepsilon=0.005$ & individual  & $1.17\pm 0.07$ & $p(x<0)=3.8\times 10^{-5}$ ($4.0\sigma$)  \\
$4<M_G<14$, no limit on $M_{\rm{tot}}$ & 1206 &  $\varepsilon=0.0025$ & individual  & $1.24\pm 0.09$ & $p(x<0)=5.0\times 10^{-5}$ ($3.9\sigma$)  \\
 $4<M_G<14$, no limit on $M_{\rm{tot}}$ & 2463 &  $\varepsilon=0.005$ & statistical  & $1.15\pm 0.06$ & $p(x<0)=1.7\times 10^{-5}$ ($4.1\sigma$)  \\
\hline
\end{tabular}
\end{center}
 \textbf{Note.}  {(1) The parameter $\gamma_{v_p}$ is the boost factor for $v_p$ (sky-projected velocity) estimated for bins with $s\ga 5$~kau. (2) $x\equiv \log_{10}(\bar{v}_{p,\rm{obs}}/\bar{v}_{p,\rm{N}})$.}
\end{table*}

\subsubsection{Auxiliary analyses} \label{sec:auxiliary}

 {In Section~\ref{sec:main}, the Newtonian analysis with $f_{\rm{multi}}=0$ showed that the observed velocities in the smaller-$s$ bins matched the Newton-predicted values while those in the larger-$s$ bins were higher than the Newton-predicted values. It is interesting to explore any possibility of attributing the boosted velocities in the larger-$s$ bins somehow to additional masses from hidden close companions or a systematically shifted magnitude-mass relation. Here I consider varying $f_{\rm{multi}}$ to make the observed velocities in the large-$s$ bins agree with the Newton-predicted values.}

\begin{figure*}
  \centering
  \includegraphics[width=0.7\linewidth]{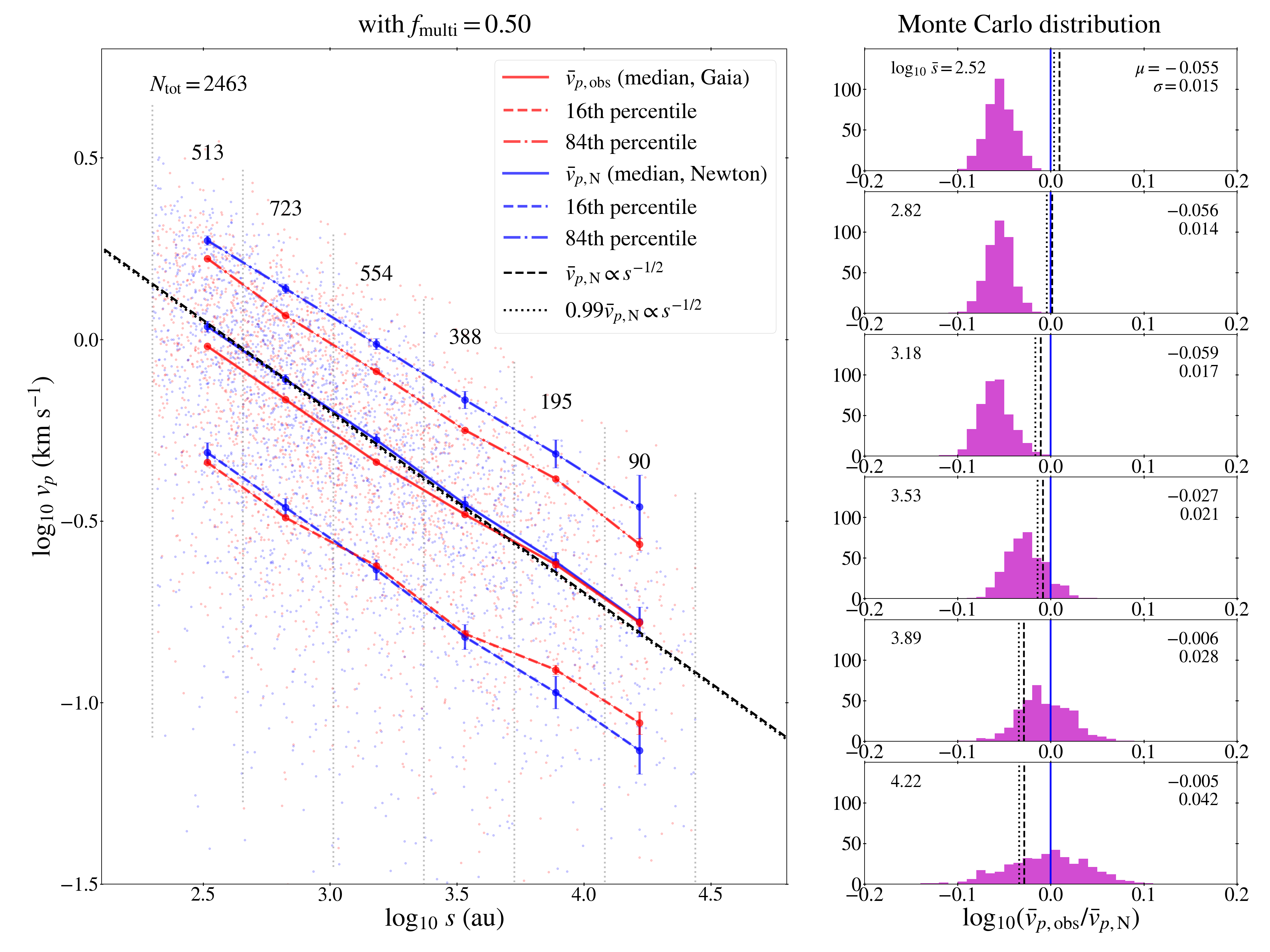}
    \vspace{-0.2truecm}
    \caption{\small 
      {The same as Figure~\ref{vp_main} but with $f_{\rm{multi}}=0.5$.}
    } 
   \label{vp_multi0_50}
\end{figure*}

 {I use the procedure of modeling close companions described in \cite{chae2023}. Figure~\ref{vp_multi0_50} shows the result with $f_{\rm{multi}}=0.5$. In this case, the observed median velocities in the two largest-$s$ bins match well the Newton-predicted medians. However, the Newton-predicted distributions are broader as revealed by the 16th and 84th percentiles due to the scatters arising from added components. More importantly, the Newton-predicted velocities in the three smallest-$s$ bins now severely deviate from the observed velocities. Thus, we would be in a Newtonian world where binaries in the three smallest-$s$ bins require $f_{\rm{multi}}\approx 0$ while those in the two largest-$s$ bins require $f_{\rm{multi}}\approx 0.5$ although two subsamples satisfy identical photometric, astrometric, and radial velocity requirements. Since this huge difference in $f_{\rm{multi}}$ between the small- and large-$s$ bins is  {ad-hoc and unlikely}, this result reinforces the results of Section~\ref{sec:main}.}

\begin{figure*}
  \centering
  \includegraphics[width=0.7\linewidth]{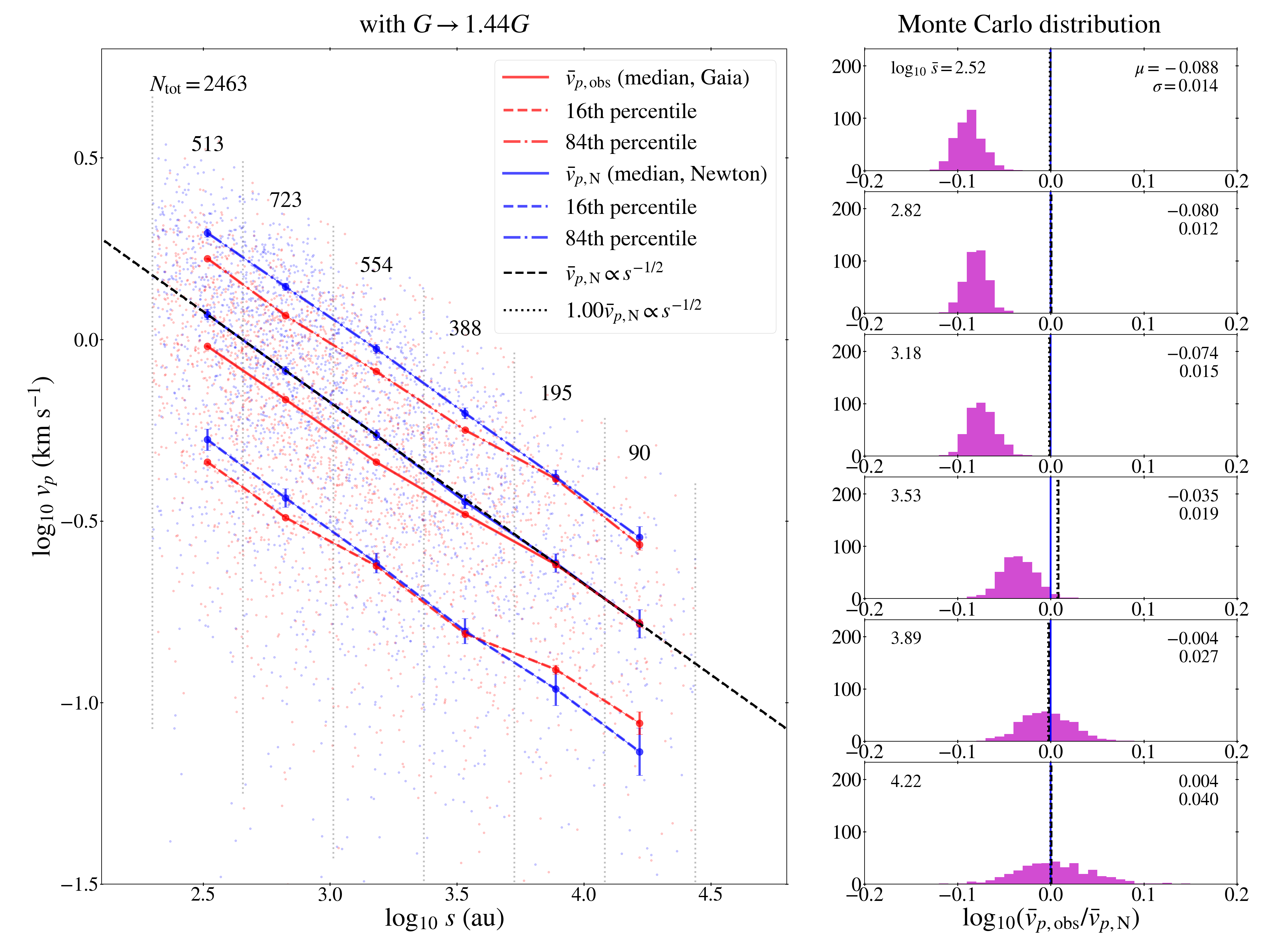}
    \vspace{-0.2truecm}
    \caption{\small 
      {The same as Figure~\ref{vp_main} but with a boosted Newton's constant $G^\prime = 1.44 G$.}
    } 
   \label{vp_Gboost}
\end{figure*}

  {I also consider a pseudo-Newtonian analysis with a rescaled Newton's constant. This analysis is motivated because the results of Section~\ref{sec:main} suggest that binaries with $s\ga 5$~kau follow pseudo-Newtonian dynamics. Figure~\ref{vp_Gboost} shows the result with $G^\prime = 1.44 G$. The observed median velocities in the two largest-$s$ bins agree well with the pseudo-Newtonian predictions.  {The observed 84th and 16th percentiles also match well the pseudo-Newtonian predictions within the statistical uncertainties.} These results suggest that binaries with $s\ga 5$~kau truly obey pseudo-Newtonian dynamics. However, the velocities in the three smallest-$s$ bins deviate severely from the pseudo-Newtonian predictions as well expected because they are in the Newtonian regime.}

\subsubsection{Physical velocity $v=\sqrt{v_p^2+v_r^2}$}  \label{sec:velprof_v}

 {Just 40 out of 2,463 binaries of the main sample have measured radial velocities satisfying the precision of $< 0.005$ required for proper motions in each dimension. These exceptional systems can be used to probe the profile of physical velocities in the 3D space.}

\begin{figure}
  \centering
  \includegraphics[width=1.\linewidth]{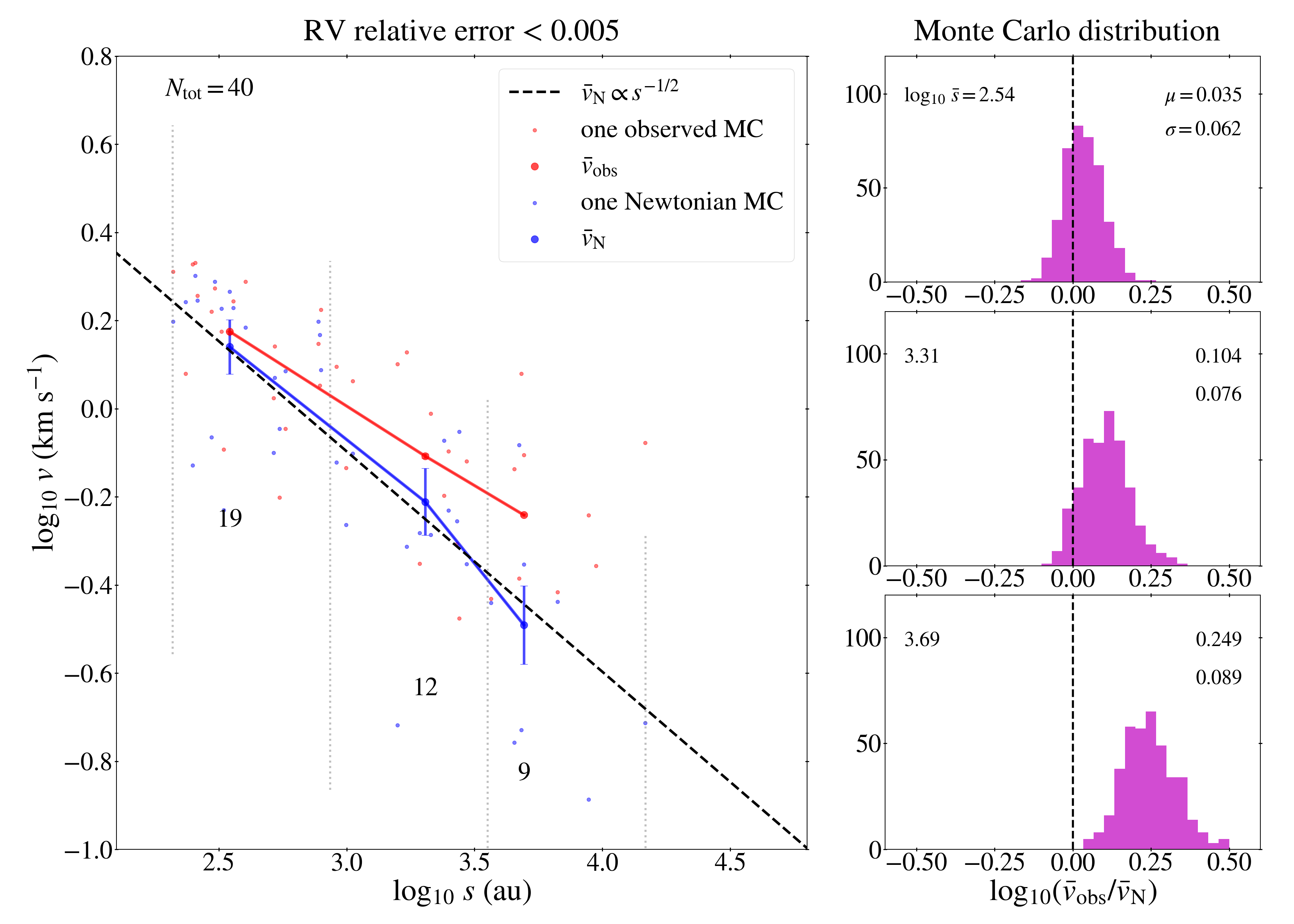}
    \vspace{-0.2truecm}
    \caption{\small 
    The result for the physical velocity $v=\sqrt{v_p^2+v_r^2}$ for binaries with exceptionally precise radial velocities. The relative errors of individual radial velocities are required to be smaller than $0.005$ as described in the text. 
    } 
   \label{v}
\end{figure}

Figure~\ref{v} shows the measured physical velocities along with the Newtonian predicted values. Only  {three} bins of $s$ are considered to obtain median velocities. As expected, the MC generated distributions of median velocities are quite broad.  {In the lowest-$s$ bin the observed median velocity matches well the Newtonian prediction. In the middle bin the observed median velocity is consistent with the Newtonian prediction within $1.5\sigma$. However, in the highest-$s$ bin satisfying $s>3.5$~kau there is an indication that $\bar{v}_{\rm{obs}}$ is higher than $\bar{v}_{\rm{N}}$ with $\log_{10}(\bar{v}_{\rm{obs}}/\bar{v}_{\rm{N}})=0.249\pm 0.089$, which represents a $2.8\sigma$ deviation.} This result is consistent with the results from the analyses of sky-projected velocity profiles. 

\section{discussion} \label{sec:discussion}

All the above results  {except for those from the auxiliary analyses of Section~\ref{sec:auxiliary}} have been obtained without any free parameters. They are derived or calculated quite naturally from the Gaia measurements, the magnitude($M_G$)-stellar mass($M$) relations from \cite{chae2023}, and the individual eccentricity ranges from \cite{hwang2022}. Are there any possibilities that any of the observational inputs  {or analyses} are grossly in error?

\begin{figure*}
  \centering
  \includegraphics[width=0.7\linewidth]{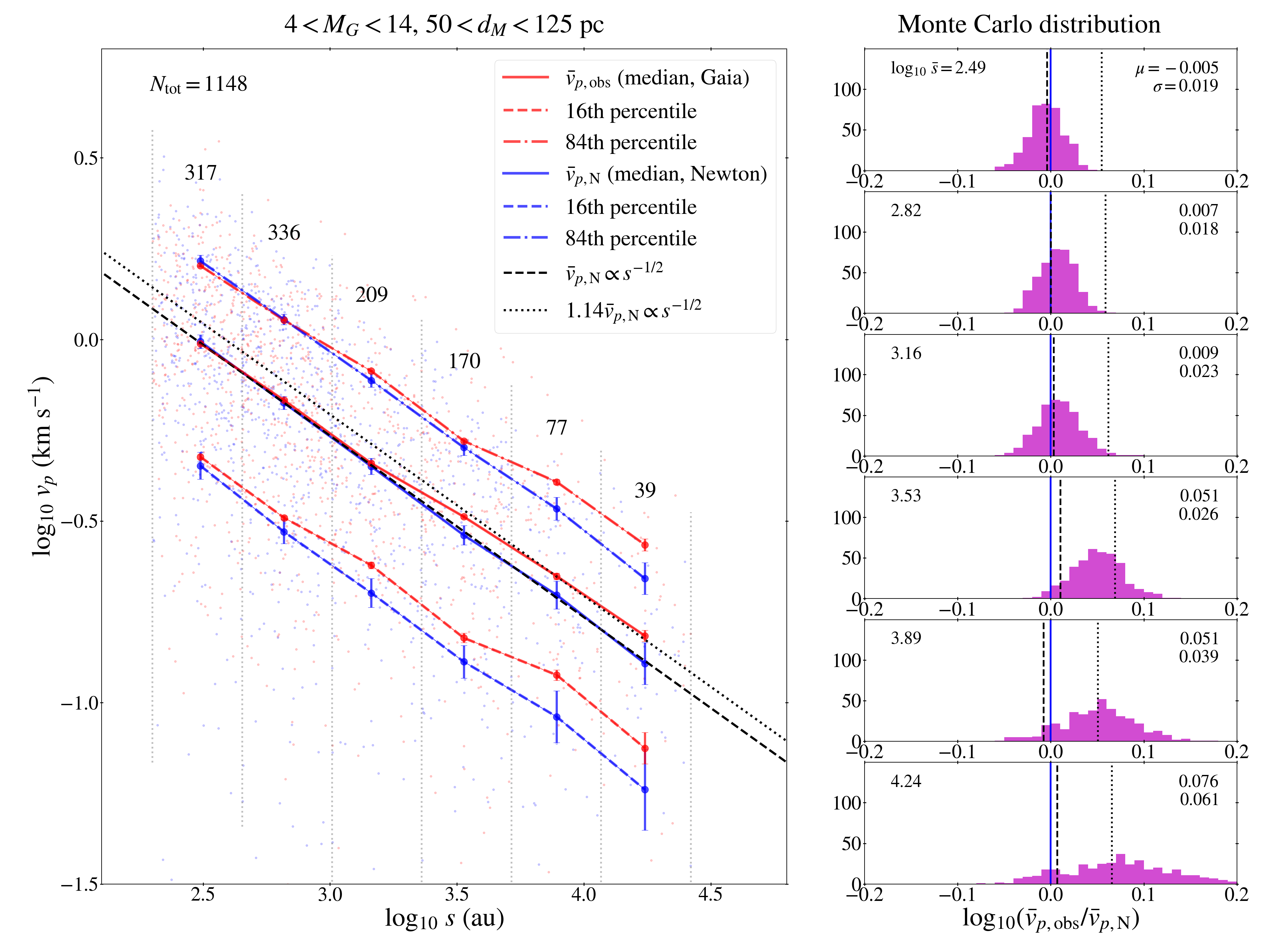}
    \vspace{-0.2truecm}
    \caption{\small 
      {The same as Figure~\ref{vp_main} but for a subsample within a narrow distance range of $50<d_M<125$~pc.}
    } 
   \label{vp_dist_50_125}
\end{figure*}

\begin{figure*}
  \centering
  \includegraphics[width=0.7\linewidth]{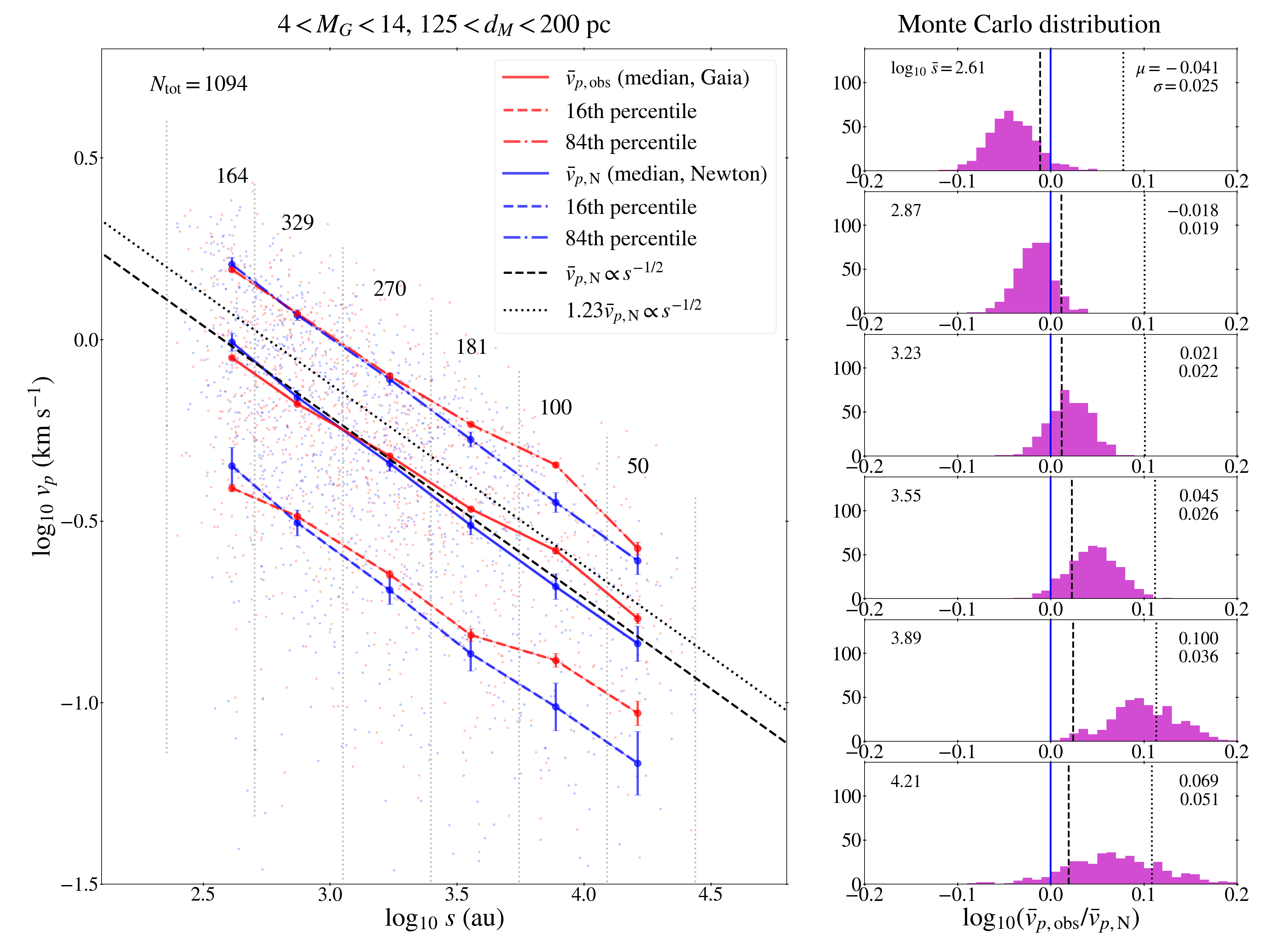}
    \vspace{-0.2truecm}
    \caption{\small 
      {The same as Figure~\ref{vp_main} but for a subsample within a narrow distance range of $125<d_M<200$~pc.}
    } 
   \label{vp_dist_125_200}
\end{figure*}

Gaia data themselves cannot be a source of systematic error because only exceptionally precise data have been used in this study  {(see Figure~\ref{verr})} and the results remain consistent as the precision increases (see Figure~\ref{delg_epsilon} and \ref{vp_eps0_0025}).  {However, one possible concern may be that the above results are based on a composite of binaries at significantly different distances ranging from 9~pc up to 200~pc. Here I consider two subsamples in significantly narrower distance ranges of $50<d_M<125$~pc and $125<d_M<200$~pc as defined in Figure~\ref{dist}. Figures~\ref{vp_dist_50_125} and \ref{vp_dist_125_200} show the results. Because sample sizes are smaller, these results have larger statistical uncertainties, but are consistent with each other and the main results. Note particularly that the $50<d_M<125$~pc sample has particularly larger statistical uncertainties because the two largest-$s$ bins have relatively few binaries.}

\begin{figure*}
  \centering
  \includegraphics[width=0.7\linewidth]{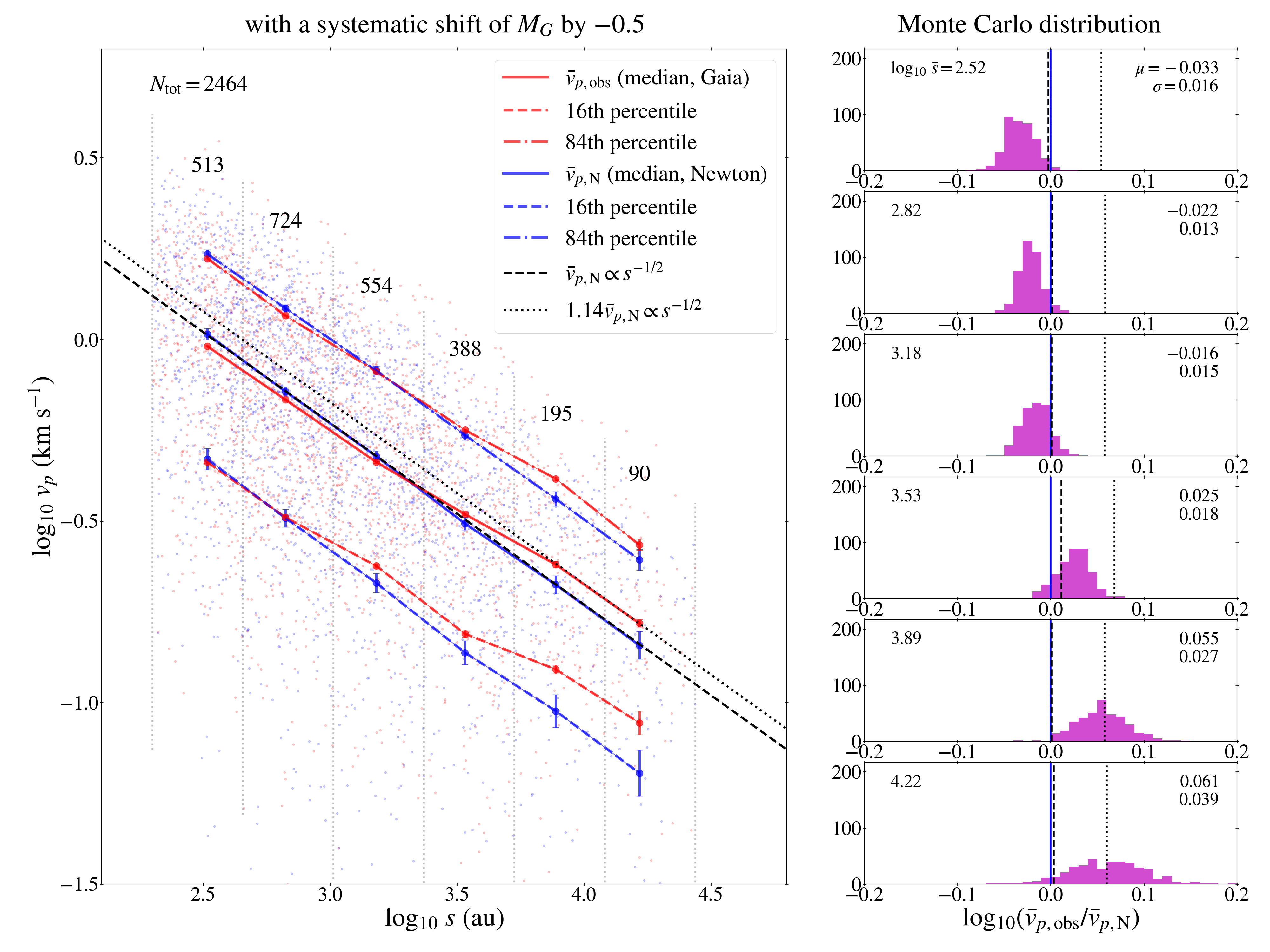}
    \vspace{-0.2truecm}
    \caption{\small 
      {The same as Figure~\ref{vp_main} but with $G$-band magnitudes systematically shifted by $-0.5$.}
    } 
   \label{vp_MG_minus}
\end{figure*}

The $M_G$-$M$ relation is also not like to be any source of systematic error. As shown in \cite{chae2023}, the two $M_G$-$M$ relations reliably cover the likely range at least for the clean magnitude range $4<M_G<14$, and the results with the two relations give consistent results (see Figure~\ref{delg_main} and compare Figure~\ref{vp_main} with Figure~\ref{vp_main_j}). Moreover, the result for the subsample with a narrower magnitude range $4<M_G<10$ (Figure~\ref{vp_narrow_massltd}) is consistent with those for the full sample. (Note that the $M_G$-$M$ relation is particularly accurate in the range $4<M_G<10$ as shown in figure~7 of \cite{chae2023}.)  {Nevertheless, I consider systematically shifting magnitudes by $\pm 0.5$~mag just as a wild possibility in the estimated $G$-band magnitudes. I note that masses of \cite{pecaut2013} agree well with the masses measured directly from confirmed close binaries at the same $K$-band magnitudes \citep{mann2019}. Thus, it suffices to consider a systematic shift in $G$-band magnitudes only. Figure~\ref{vp_MG_minus} shows the result with a shift of $-0.5$. Even in this case, the median velocities in the two largest-$s$ bins deviate significantly with a velocity boost factor of $\approx 1.14$ while the two smallest-$s$ bins deviate in the opposite direction. }

  Kinematic contaminants such as hidden close companions cannot be a source of systematic error. Because binaries with small and large $s$ values have similar total masses and satisfy the same selection criteria, if additional masses should be added to the binaries, similar masses must be added to the binaries regardless of their separations. Then, the fractional boost in velocity is similar for all binaries regardless of $s$ because the boosted-to-initial velocity ratio $v^\prime/v = \sqrt{M^\prime/M}$ is independent of $s$. As shown in Figure~\ref{vp_multi0_50}, when additional masses are controlled by $f_{\rm{multi}}$, the two largest-$s$ bins require $f_{\rm{multi}}=0.5$ to match Newtonian dynamics while the three smallest-$s$ bins require $f_{\rm{multi}}=0$. This  {extremely unlikely} difference means that it  {seems} impossible to attribute the gravitational anomaly to differential kinematic contaminants of hidden close companions.  {This view is bolstered by the fitted values of $f_{\rm{multi}}$ in subsamples with progressively stricter kinematic criteria as presented in Appendix~\ref{sec:kincut}. When kinematic criteria get stricter and stricter, $f_{\rm{multi}}$ gets lower and lower, approaching zero eventually. Thus, it is hardly expected to be as high as $f_{\rm{multi}}=0.5$ in the statistically pure binary sample selected in this work. Because the sample of 2,463 ``pure'' wide binaries is made public, this claim can be directly tested with observations (see \citealt{manchanda2023}). }
  
\begin{figure*}
  \centering
  \includegraphics[width=0.7\linewidth]{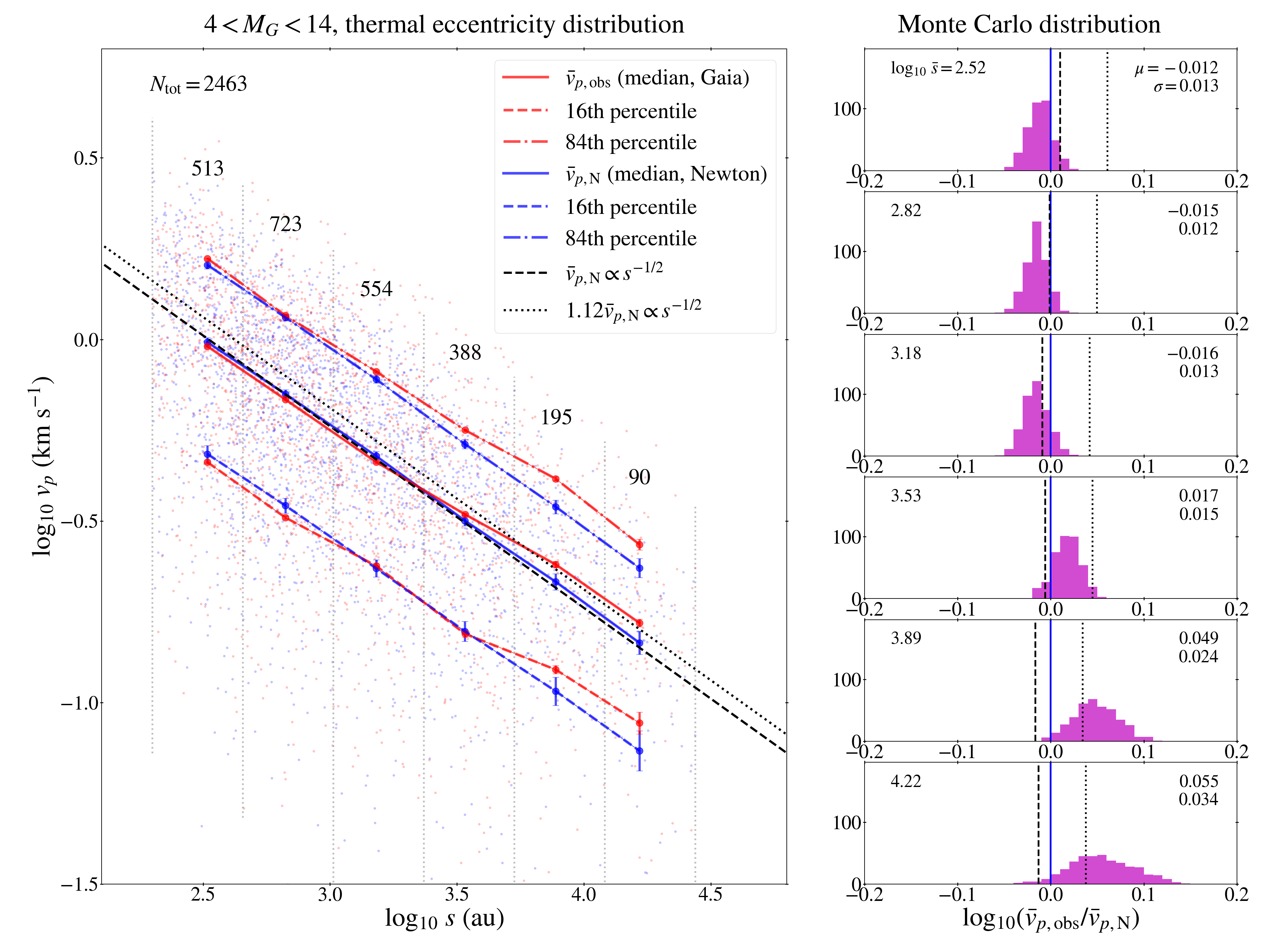}
    \vspace{-0.2truecm}
    \caption{\small 
    The same as Figure~\ref{vp_main} but with eccentricities from a `thermal' probability distribution $p(e)=2e$ ignoring individual eccentricities. The result is deliberately biased toward a weakened boost of velocities. Nevertheless, the boost is still very significant.
    } 
   \label{vp_main_thermal}
\end{figure*}

The individual eccentricity ranges from \cite{hwang2022} are the best available empirical information on eccentricities at present.  {I have also considered the power-law distribution of eccentricities with $s$-dependent exponent (Equation~\ref{eq:alpha}).} However, here I consider the uniform thermal probability distribution of eccentricity $p(e)=2e$ for all binaries to gauge possible source of systematic error arising from eccentricities. Note that this choice is deliberately biased from the empirical information (see figure~24 of \cite{chae2023}). Figure~\ref{vp_main_thermal} shows the result with the thermal eccentricity distribution. In this case, the deviation is somewhat weakened but the combined statistical significance of the deviations in the  {two largest-$s$} bins is $3.7\sigma$. Thus, it is not possible to do away with the deviations by reasonably modifying eccentricities. 

\section{Conclusion and future prospects} \label{sec:conclusion}

A sample of  {2,463} \emph{statistically} pure wide binaries in the mass range $0.5\la M_{\rm{tot}}/ {\rm{M}}_\odot \la 2$ provides two crucial results on non-relativistic gravitational dynamics.

First, the observed orbital motions of binaries with relatively small sky-projected separations ($s\la 2$~kau) are statistically consistent with Newtonian dynamics, naturally and without any adjustment of Gaia data or other observational inputs. This is a nontrivial result that provides direct evidence that currently available gravity theories including Newtonian and Milgromian theories hold in the non-relativistic regime of accelerations $\ga 10^{-8}$~m~s$^{-2}$.

Second, the observed orbital motions of binaries with relatively larger $s\ga 2$~kau are statistically \emph{inconsistent} with Newtonian dynamics. The Gaia measured sky-projected velocities are boosted with a statistical significance of $\approx 5.0\sigma$. In the bin of the investigated  {range $5<s<30$~kau, the velocity boost factor is measured to be $\gamma_{v_p} = 1.20\pm 0.06$ (stat) $\pm 0.05$ (sys) (see Table~\ref{tab:main_result}).} When the pure binaries are analyzed in the acceleration plane defined in \cite{chae2023}, the kinematic acceleration $v^2/r$ with MC-deprojected $v$ and $r$ is systematically higher in the low-acceleration regime $\la 10^{-9}$~m~s$^{-2}$ than the corresponding Newtonian prediction, with an acceleration boost factor of $\gamma_g=1.49^{+0.21}_{-0.19}$ satisfying the expected relation $\gamma_g = \gamma_{v_p}^2$.

In a small sample of  {40} pure wide binaries with exceptionally precise Gaia radial velocities, the physical velocity ($v$) is directly measured (i.e. without any deprojection). Despite the small number statistics, it is seen that the mean measured velocity in the largest-$s$ bin is boosted compared with the Newtonian prediction while that in the smallest-$s$ bin naturally matches the Newtonian prediction.

The present results from analyses of statistically pure binaries provide a robust confirmation of the results of \cite{chae2023} for a much larger general sample that includes hierarchical systems with undetected companions. The present results also complement \cite{hernandez2023} who obtained a similar boost in projected velocities but with a lower or roughly quantified statistical significance.

Just recently and almost concurrently with this work, \cite{hernandez2023a} has carried out a statistical analysis of the \cite{hernandez2023} sample of 667 wide binaries within 125~pc. They obtained a boost factor of $\gamma_g=1.512\pm 0.199$ which is in good agreement with the results from this work and \cite{chae2023}.

 {Unlike this work and other recent studies \citep{chae2023,hernandez2023,hernandez2023a}, \cite{banik2023} claimed an opposite conclusion and argued particularly that the gravitational anomaly obtained by \cite{chae2023} was largely affected by kinematic contaminants. As shown in Figure~\ref{verr}, the sample of pure binaries used in this work already satisfies the \cite{banik2023} kinematic cut and yet shows essentially the same gravitational anomaly reported by \cite{chae2023,hernandez2023,hernandez2023a}. Moreover, as presented in Appendix~\ref{sec:kincut}, new results with kinematic cuts imposed confirm the gravitational anomaly.}

The evidence for the gravity boost in the low acceleration regime is now clear enough although the scientific community should keep gathering further evidence from future observations. What seems now more important is to precisely characterize the gravity boost to the point that the theoretical direction can be narrowed down. Given that precise radial velocities measured in just {40} binaries already show a mild evidence of the gravity boost in the low-acceleration regime (Figure~\ref{v}), precise measurements of radial velocities in more pure binaries in the future may turn out to be quite fruitful in characterizing the gravity in the low acceleration regime. 

 {In principle, theoretical interpretations of the gravitational anomaly obtained here and in \cite{chae2023} and \cite{hernandez2023} are wide open. However, the most straightforward interpretation at hand is that nonrelativistic gravitational dynamics is governed/described by MOND-type Lagrangian theories of gravity \citep{bekenstein1984,milgrom2010,milgrom2023}.} However, because MOND-type Lagrangian theories are nonrelativistic phenomenological theories, something like phenomenological quantization rules before the full development of quantum physics, it is unclear what will be the underlying fundamental theory that will explain the MOND phenomenology eventually. Because MOND breaks the strong equivalence principle \citep{chae2020b,chae2021,chae2022c} while keeping the Einstein equivalence principle, even non-quantum gravity must be different from Einstein's general relativity \citep{einstein1916} and reformulated  {encompassing the successful aspects of both MOND and general relativity} perhaps in the spirit of Mach's principle. 

Hypothetical dark matter was introduced as a solution to gravitational anomalies in the presently investigated low-acceleration regime in galaxies and galaxy clusters assuming that standard gravity holds in that regime. Now that standard gravity breaks down in the same low-acceleration regime regardless of hypothetical dark matter and in agreement with MOND,  {dark matter interpretation is seriously questioned as} a valid solution. Thus, no direct detection of dark matter despite intensive worldwide campaigns can now be seen as a natural outcome. Because there has been no direct detection of dark matter, all circumstantial arguments and indirect ``evidence'' for dark matter assuming standard gravity  {can} now be overridden by the present direct evidence for the breakdown of standard gravity. This means that the dark matter paradigm  {seems} now doomed to be abandoned and we are entering an era of a paradigm shift. Implications of the gravitational anomaly in the low-acceleration regime for astrophysics, cosmology, and fundamental physics are truly far-reaching.

 {In particular, the standard cosmology based on general relativity  {seems} no longer valid even in principle. Development of MOND-based cosmology and structure formation (e.g., \citealt{sanders1998,sanders2001,sanders2008,wittenburg2023}) is now well-motivated and much-needed in parallel with theoretical advancement of MOND (e.g.\ \citealt{thomas2023}).}

\section*{Acknowledgments}
The author thanks Kareem El-Badry for discussion on radial velocities of stars.  {The revised version was based on an insightful report for which the author thanks the referee, a plenary talk given at the Korean Astronomical Society Fall 2023 meeting, and an invited talk given at the Pacific Rim Conference on Stellar Astrophysics held at Sejong University in October 2023. In particular, the author spotted an error in the code described in \cite{chae2023}, which is typographic in nature, while preparing for the plenary talk, and the correction was reflected in this revision.} This work was supported by the National Research Foundation of Korea (grant No. NRF-2022R1A2C1092306). 

\bibliographystyle{aasjournal}

\newpage

\appendix

\section{A correction to Chae (2023)} \label{sec:correction}

 {In a subsection of \cite{chae2023} I inadvertently used a relation that was valid only for circular orbits. This error occurred in the subsection ``3.4 \emph{Newtonian Simulation}'' of \cite{chae2023} while the correct relation was used in the subsection ``3.1 \emph{Monte Carlo deprojection of the observed 2D motion to the 3D motion}''. This error affected mainly the calibrated/fitted values of $f_{\rm{multi}}$ in matching observed accelerations with Newton-predicted accelerations at a high acceleration bin.  Here I correct the relation and update representative results of \cite{chae2023}. The published codes are also updated.}

 {For the geometry of Figure~\ref{orbit} the sky-projected relative velocity components to replace equation~(18) of \cite{chae2023} are given by
  \begin{equation}
    \begin{array}{lll}
    v_{p,x^\prime} & = & v(r) \cos \psi, \\
    v_{p,y^\prime} & = & v(r) \cos i \sin\psi, \\
    \end{array}
    \label{eq:vpcomp}
  \end{equation}
  where $v(r)$ and $\psi$ are given by Equations~(\ref{eq:vN}) and (\ref{eq:psi}).}

 {Tests with Newtonian mock data based on the revised code indicate that it is more suitable to compare the binned medians of Newtonian Monte Carlo (MC) velocities with the binned medians of the most likely values without scatters of proper motions from one MC to another. However, this distinction becomes less unimportant when only high-precision proper motions (PMs) are used.}

 {With the correction the fitted values of $f_{\rm{multi}}$ are now $f_{\rm{multi}}=0.48$ for the main sample of 26,615 wide binaries within 200~pc and $0.36$ for the subsample of 19,716 wide binaries with PM relative errors $<0.005$. The former value is significantly smaller than $0.65$ reported in \cite{chae2023}. Because $>1''$ resolved hierarchical systems were excluded in the binary samples that were taken from \cite{elbadry2021}, the lowered values of $f_{\rm{multi}}$ are now more compatible with the current observational range of $0.2\la f_{\rm{multi}} \la 0.5$.}

\begin{figure*}
  \centering
  \includegraphics[width=0.8\linewidth]{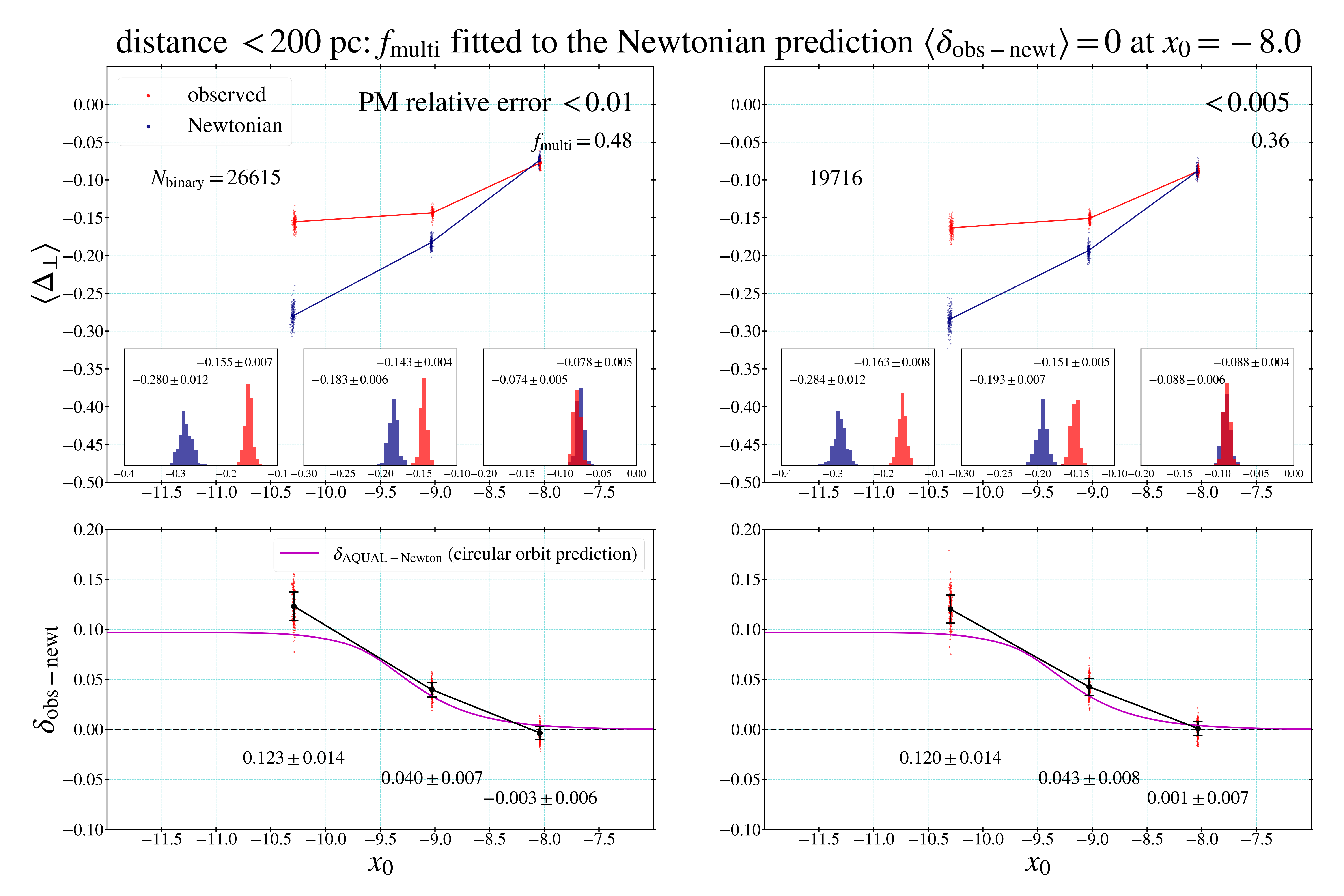}
    \vspace{-0.2truecm}
    \caption{\small 
     { Gravitational anomalies of \cite{chae2023} are revised for the main samples with a correction to the published code with Equation~(\ref{eq:vpcomp}). The upper panels show distributions of median orthogonal deviations $\langle\Delta_\bot\rangle$ (defined in Figure~\ref{RAR_main}) from 200 MC results. The upper right panel is with the $J$-band-based $M_G$-$M$ relation. The bottom panels show distributions of the difference $\delta_{\rm{obs-newt}}\equiv \langle\Delta_\bot\rangle_{\rm{obs}} - \langle\Delta_\bot\rangle_{\rm{N}}$. Parameter $f_{\rm{multi}}$ was fitted through an iterative procedure so that $\delta_{\rm{obs-newt}} = 0$ is satisfied at $x_0\approx -8.0$. The magenta curve in the bottom panels represents the AQUAL prediction for circular orbits with the Milky Way external field.}
    } 
   \label{delg_new}
\end{figure*}

 {Figure~\ref{delg_new} shows the revised results for the main samples with relative PM error $<0.01$ or $0.005$ based on the standard inputs including the individual eccentricities. The inferred gravitational anomalies are now somewhat stronger with higher statistical significance. In either case, the lowest acceleration bin shows a $>8\sigma$ deviation while the middle bin shows a $>5\sigma$ deviation. The acceleration boost factor (Equation~\ref{eq:gamma_g}) is given by $\gamma_g=1.49\pm 0.07$ (for the larger sample) or $\gamma_g=1.48\pm 0.07$ (for the smaller sample). These values are compatible with the AQUAL prediction of $\approx 1.4$ for circular orbits.}

\begin{figure*}
  \centering
  \includegraphics[width=0.8\linewidth]{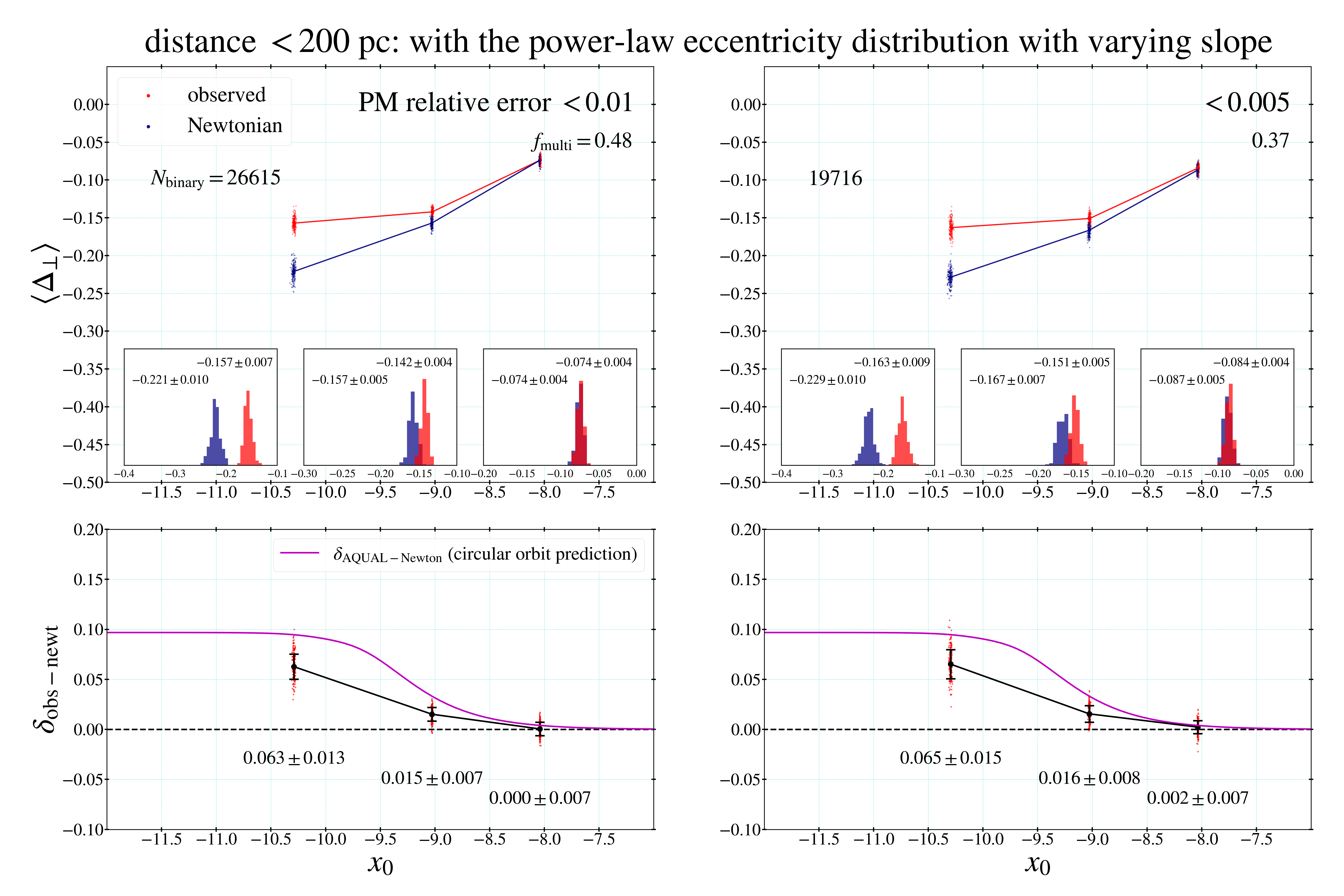}
    \vspace{-0.2truecm}
    \caption{\small 
     { Same as Figure~\ref{delg_new} but with eccentricities drawn from the power-eccentricity distribution with the slope varying with $s$ as given by Equation~(\ref{eq:alpha}).}
    } 
   \label{delg_new_systematice}
\end{figure*}

 {Figure~\ref{delg_new_systematice} shows results with eccentricities drawn from a systematically varying statistical distribution rather than individual eccentricities from \cite{hwang2022}. Note that the statistical distribution is also provided by \cite{hwang2022}. With these eccentricities the gravitational anomalies are weaker but still very significant from the lowest acceleration bin alone. The difference between Figure~\ref{delg_new} and Figure~\ref{delg_new_systematice} highlights the importance of eccentricities in precisely characterizing the revealed gravitational anomaly.}

\section{Effects of kinematic cuts in sampling on wide binary tests of gravity} \label{sec:kincut}

 {Recently, \cite{banik2023} have emphasized the potential importance of kinematic cuts in sample selection for wide binary tests of gravity.  The sample of pure binaries from the present study already satisfies their kinematic requirements (see Figure~\ref{verr}). Here I test the main sample of \cite{chae2023} and its subsamples with some kinematic cuts including that suggested by \cite{banik2023}.}

 {I consider two quantities from kinematic data. One is the estimated uncertainty of $\tilde{v}$ given by Equation~(\ref{eq:vtilde}). This is similar to that suggested by \cite{banik2023} differing only slightly in the way that the uncertainty of the Newtonian circular velocity $v_c$ is accounted for. \cite{banik2023} suggest that }
\begin{equation}
  \sigma_{\tilde{v}} < \beta\max{(1,\tilde{v}/2)}\text{ with }\beta=0.1
  \label{eq:vtildecut}
\end{equation}
  {may provide a good compromise between the quality and quantity of data. I note that this choice is somewhat arbitrary and they did not consider the effects of varying $\beta$ or the criterion itself. Moreover, the parameter $\tilde{v}$ is a ratio of quantities, both of which include projected quantities. }

  {Because wide binary test is intended to test whether the observed velocity or velocity squared over radius (i.e. kinematic acceleration) is boosted or not compared with the Newtonian prediction, I also consider a more direct cut based on the uncertainty of the observed sky-projected velocity $v_p$.  }
 
\begin{figure*}
  \centering
  \includegraphics[width=0.7\linewidth]{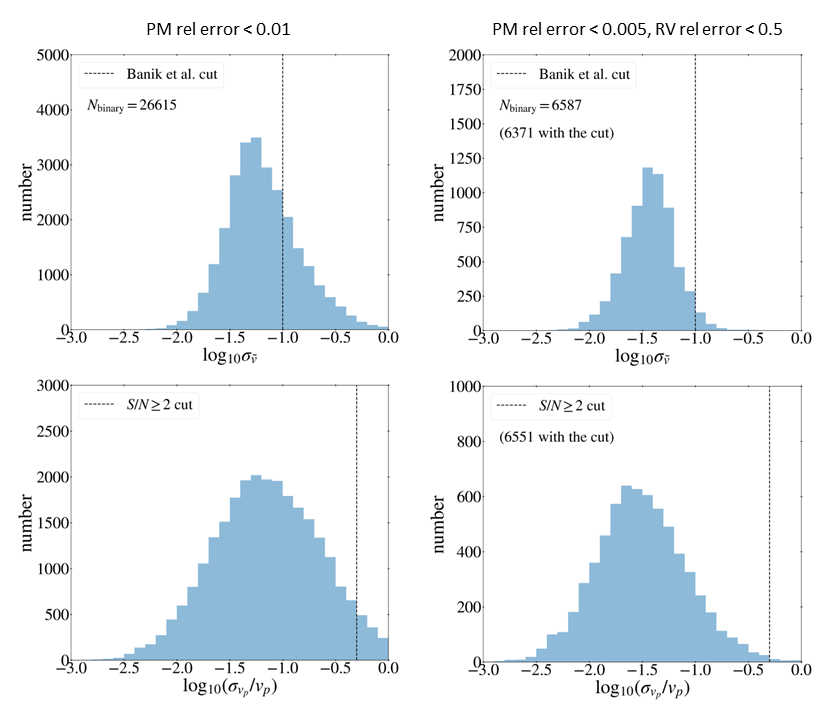}
    \vspace{-0.2truecm}
    \caption{\small 
      {The left column shows the distributions of two uncertainties $\sigma_{\tilde{v}}$ and $\sigma_{v_p}/v_p$ for the main sample of \cite{chae2023}. The right column is the same for a subsample selected with a tighter precision on PM and additional requirements on radial velocities. In the latter, I require that radial velocities of both components have $S/N>2$ and match each other within $3\sigma$. Note that the subsample largely satisfies the kinematic cuts indicated by vertical dashed lines.}
    } 
   \label{verr2}
\end{figure*} 

 {The left column of Figure~\ref{verr2} shows the distributions of the two uncertainty parameters for the main sample of \cite{chae2023} with the PM error cut $<0.01$. About 27\% of binaries from the \cite{chae2023} main sample do not pass the cut defined by Equation~(\ref{eq:vtildecut}) (this is higher than 21\% obtained by \cite{banik2023} because my cut is somewhat stricter). For the \cite{chae2023} sample with the stricter PM error cut $<0.005$, a smaller fraction of about 19\% do not pass the cut.}

 {But, even if the cut such as Equation~(\ref{eq:vtildecut}) can be fully justified, what effect would it have on inferring gravity from wide binaries? \cite{banik2023} argued that a qualitative examination of the scaling of $\tilde{v}$ with sky-projected radius normalized by the MOND radius $r_M\equiv \sqrt{GM_{\rm{tot}}/a_0}$ could question the conclusion of \cite{chae2023}. However, the question can only be answered through detailed quantitative analyses. Also, the \cite{banik2023} cut itself needs to be examined. Thus, a thorough investigation could be done properly only through a heavy independent work. Here I present only the default results based on the cut given by Equation~(\ref{eq:vtildecut}) and the standard inputs of \cite{chae2023}.}

\begin{figure*}
  \centering
  \includegraphics[width=0.8\linewidth]{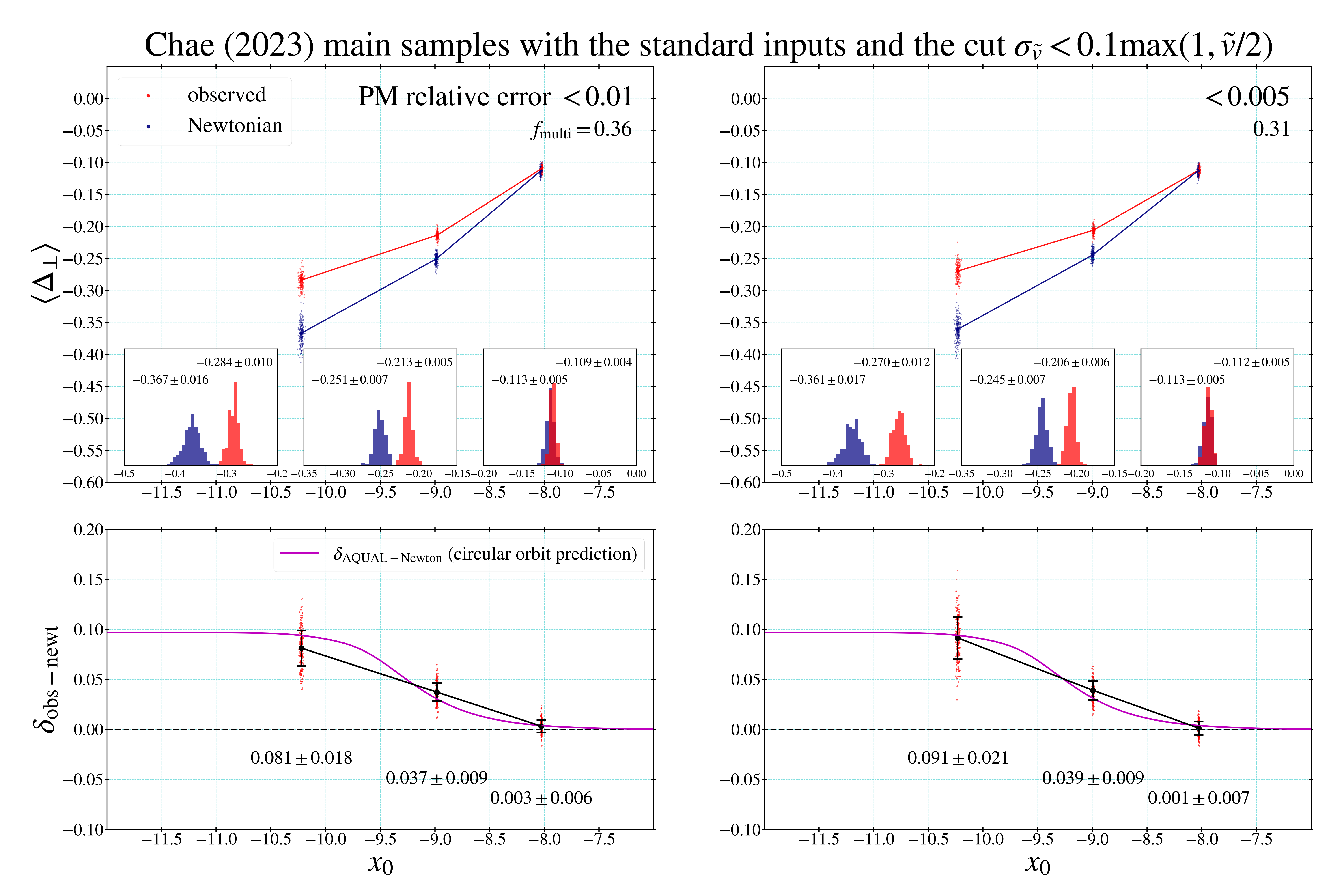}
    \vspace{-0.2truecm}
    \caption{\small 
      { Same as Figure~\ref{delg_new} but with the kinematic constraint of Equation~(\ref{eq:vtildecut}) imposed.}
    } 
   \label{delg_main_vtilerr}
\end{figure*}

 {Figure~\ref{delg_main_vtilerr} shows the results for the two main samples of \cite{chae2023}. The deviations from the corresponding Newtonian predictions are only mildly lowered compared with those without the kinematic constraint (Figure~\ref{delg_new}). In particular, the result (the right panel) for the sample with PM error $<0.005$ is barely affected by the cut. These results indicate that the \cite{banik2023} argument based on the qualitative appearances of $\tilde{v}$ curves does not hold.  However, these results are not surprising because much more precise data used in the main part of this work already agreed with the results of \cite{chae2023}.}

 {Figure~\ref{delg_main_vtilerr} also shows that the fitted values of $f_{\rm{multi}}$ are lowered if the kinematic constraint of Equation~(\ref{eq:vtildecut}) is imposed. This is easily understood by the fact that hidden close companions are likely to add some kinematic noise. The fitted values of $f_{\rm{multi}}=0.36$ and $0.31$ are quite reasonable considering that some noisy data were removed. Although the sample of \cite{banik2023} is not the same as the samples used here, the extremely high value of $f_{\rm{multi}}\approx 0.70$ for their preferred Newtonian model indicates that their preference of Newtonian gravity is likely to be due to the largely biased $f_{\rm{multi}}$. The bias may have occurred because they chose not to calibrate $f_{\rm{multi}}$ using Newtonian regime data.  }

 {I also consider a more reliable subsample than the main samples shown in Figure~\ref{delg_main_vtilerr}. In this sample, it is required that $S/N$ values of the reported radial velocities for both components are greater than 2 and two radial velocities match each other within $3\sigma$ (adjusting Equation~(\ref{eq:delvr})). This sample will be intermediate between the main samples of \cite{chae2023} and the most precise pure binary sample from this work. The right column of Figure~\ref{verr2} shows the distributions of the two uncertainty parameters for this sample. Interestingly, both kinematic constraints are largely satisfied by this sample. I carry out detailed tests of gravity with this sample. }

\begin{figure*}
  \centering
  \includegraphics[width=0.8\linewidth]{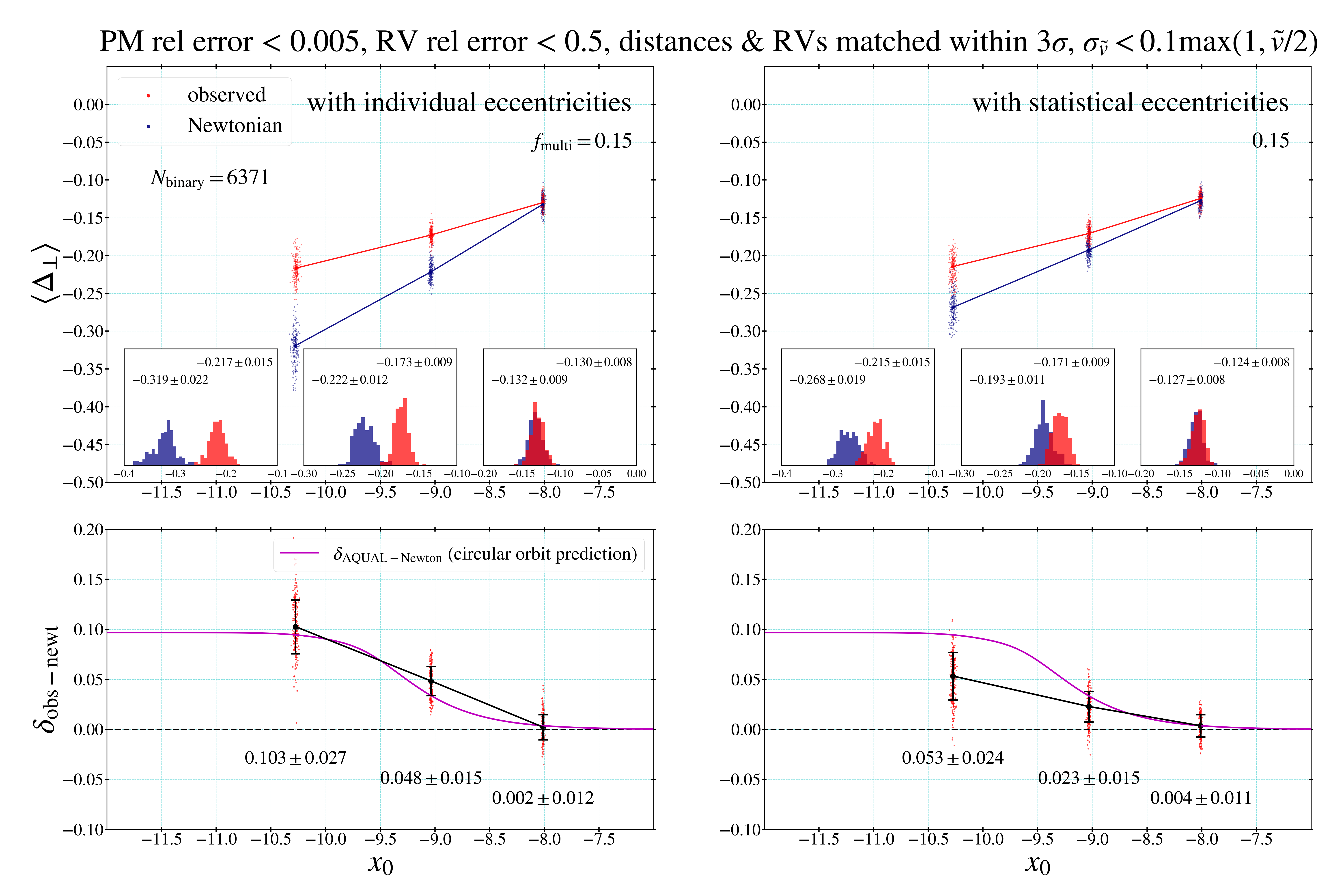}
    \vspace{-0.2truecm}
    \caption{\small 
      {Similar to the right panel of Figure~\ref{delg_main_vtilerr} but for a subsample with radial velocities shown in the upper-right panel of Figure~\ref{verr2}. Note that statistical eccentricities are also considered to provide systematically biased results. }
    } 
   \label{delg_vtilerr}
\end{figure*}

\begin{figure*}
  \centering
  \includegraphics[width=0.8\linewidth]{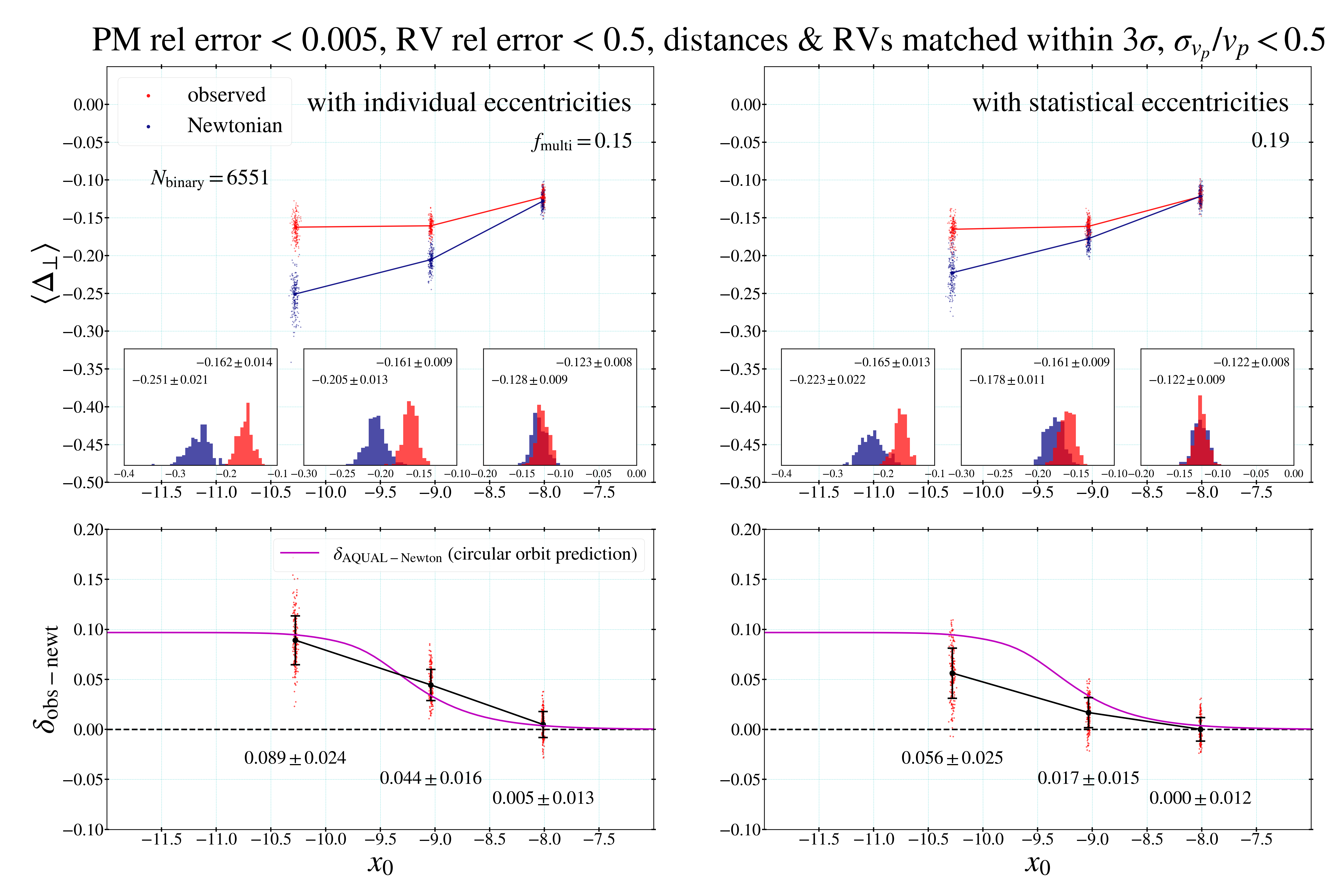}
    \vspace{-0.2truecm}
    \caption{\small 
      {The same as Figure~\ref{delg_vtilerr} but with a different kinematic cut shown in the lower-right panel of Figure~\ref{verr2}. }
    } 
   \label{delg_verr}
\end{figure*}

 {Figures~\ref{delg_vtilerr} and \ref{delg_verr} show the results from testing gravity with the subsamples shown in the right column of Figure~\ref{verr2}. Two samples with different kinematic cuts imposed return similar results. The results with individual eccentricities match the AQUAL prediction remarkably well. It is also interesting to note that for these relatively more reliable samples (compared with the \cite{chae2023} main samples) there is a clear indication for the gravitational anomaly, though somewhat weakened, even with statistical eccentricities that are less informative than individual eccentricities. The fitted values of $f_{\rm{multi}}=0.15$ or $0.19$ are even lower than those shown in Figure~\ref{delg_main_vtilerr}, indicating further that the \cite{banik2023} value of $\approx 0.70$ is unlikely for samples satisfying kinematic cuts. }   

\begin{figure*}
  \centering
  \includegraphics[width=0.7\linewidth]{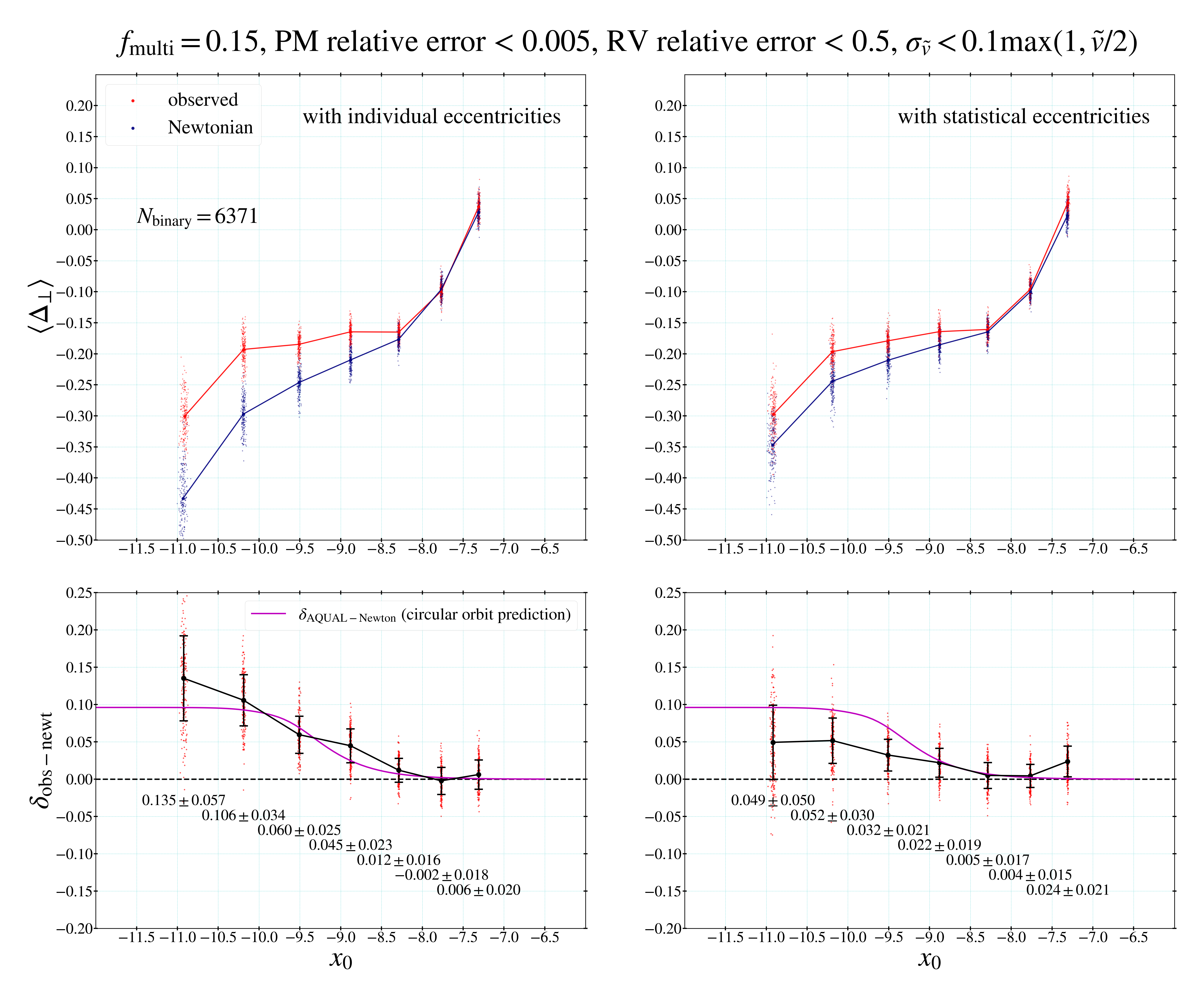}
    \vspace{-0.2truecm}
    \caption{\small 
      {Similar to Figure~\ref{delg_vtilerr} but for finer bins. }
    } 
   \label{delg_vtilerr_7bins}
\end{figure*}

\begin{figure*}
  \centering
  \includegraphics[width=0.7\linewidth]{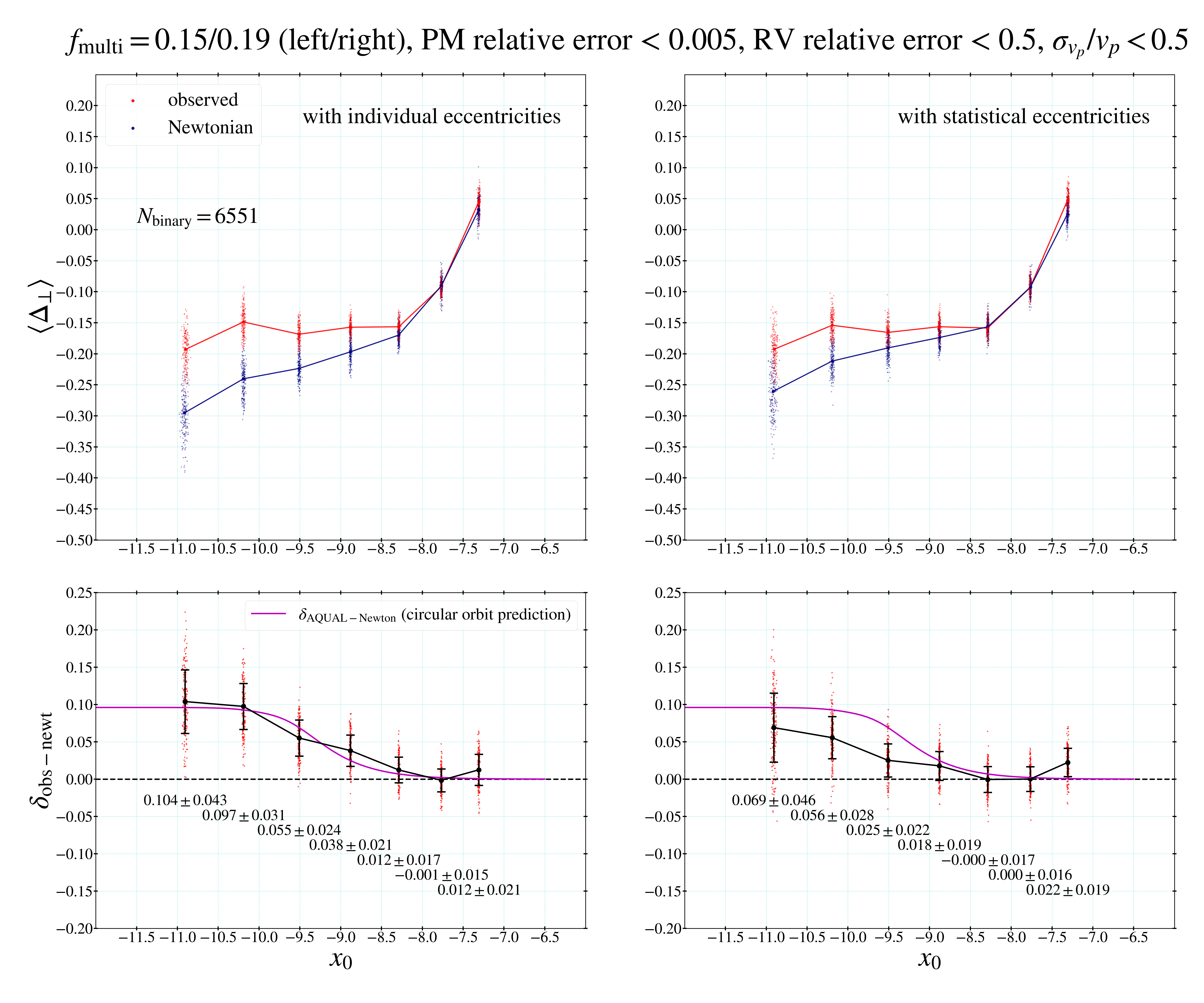}
    \vspace{-0.2truecm}
    \caption{\small 
      {The same as Figure~\ref{delg_verr} but for finer bins. }
    } 
   \label{delg_verr_7bins}
\end{figure*}

 {Finally, Figures~\ref{delg_vtilerr_7bins} and \ref{delg_verr_7bins} show the results for seven bins from testing gravity with the subsamples used for Figures~\ref{delg_vtilerr} and \ref{delg_verr}. The results with the more reliable individual eccentricities closely follow the AQUAL prediction for the entire range of acceleration covered by the data. }

\end{document}